\documentclass[reprint,prd,twocolumn,superscriptaddress,nofootinbib,longbibliography]{revtex4-1}

\usepackage{color}
\usepackage{amsmath}
\usepackage{graphicx}
\usepackage{scalefnt}
\usepackage{hyperref}

\usepackage{grffile} 

\newcommand{\lt}{\left}
\newcommand{\rt}{\right}

\newcommand{\ov}{\overline}
\newcommand{\eq}[1]{Eq.~(\ref{#1})}

\newcommand{\gev}{\,\mbox{GeV}}
\newcommand{\tev}{\,\mbox{TeV}}

\newcommand{\Bbar}{\bar{B}}

\newcommand{\bb}{\ensuremath{B\!-\!\Bbar\,}}

\newcommand{\bbq}{\ensuremath{B_q\!-\!\Bbar{}_q\,}}

\newcommand{\lqcd}{\Lambda_{\rm QCD}} 
\newcommand{\dm}{\ensuremath{\Delta M}}
\newcommand{\dg}{\ensuremath{\Delta \Gamma}}

\newcommand{\ket}[1]{| #1 \rangle }

\graphicspath{ {figs/} }

\allowdisplaybreaks

\begin{document}

\hfill TTP22-030,  P3H-22-051

\title{The width difference in the  $\bb$ system at
  next-to-next-to-leading order of QCD}
\author{Marvin Gerlach}
\email{gerlach.marvin@protonmail.com}
\affiliation{Institut f\"ur Theoretische Teilchenphysik,
  Karlsruhe Institute of Technology (KIT), 76128 Karlsruhe, Germany}
\author{Ulrich Nierste}
\email{ulrich.nierste@kit.edu}
\affiliation{Institut f\"ur Theoretische Teilchenphysik,
  Karlsruhe Institute of Technology (KIT), 76128 Karlsruhe, Germany}
\author{Vladyslav Shtabovenko}
\email{v.shtabovenko@kit.edu}
\affiliation{Institut f\"ur Theoretische Teilchenphysik,
  Karlsruhe Institute of Technology (KIT), 76128 Karlsruhe, Germany}
\author{Matthias Steinhauser}
\email{matthias.steinhauser@kit.edu}
\affiliation{Institut f\"ur Theoretische Teilchenphysik,
  Karlsruhe Institute of Technology (KIT), 76128 Karlsruhe, Germany}

\begin{abstract}

  We extend the theoretical prediction for the width difference
  $\Delta \Gamma_q$ in the mixing of neutral $B$ mesons in the Standard Model
  to next-to-next-to-leading order in $\alpha_s$. To this aim we calculate
  three-loop diagrams with two $|\Delta B|=1$ current-current operators
  analytically.  In the matching between $|\Delta B|=1$ and $|\Delta B|=2$
  effective theories we regularize the infrared divergences dimensionally and
  take into account all relevant evanescent operators.  Further elements of
  the calculation are the two-loop renormalization matrix $Z_{ij}$ for the
  $|\Delta B|=2$ operators and the $\mathcal{O}(\alpha_s^2)$ corrections to
  the finite renormalization that ensures the $1/m_b$ suppression of the
  operator $R_0$ at two-loop order.  Our theoretical prediction reads
  $\Delta\Gamma_s/\Delta M_s = {(4.33\pm 0.93)\cdot 10^{-3}}$ if expressed in
  terms of the bottom mass in the $\overline{\rm MS}$ scheme and
  $\Delta\Gamma_s/\Delta M_s = {(4.20\pm 0.95)\cdot 10^{-3}}$ for the use of the
  potential-subtracted mass. While the controversy on $|V_{cb}|$ affects both
  $\Delta\Gamma_s$ and $\Delta M_s$, the ratio $\Delta\Gamma_s/\Delta M_s$ is
  not affected by the uncertainty in $|V_{cb}|$ .
\end{abstract}

\maketitle



\bigskip {\bf Introduction.}  The weak interaction of the Standard Model
(SM) permits transitions between a neutral $B_q$ meson and its
antiparticle $\Bbar_q$, where $q=d$ or $s$. The corresponding transition
amplitude is mediated by box diagrams with $W$ bosons and up-type quarks
$u$, $c$, or $t$ on the internal lines.  The time evolution of the
two-state system $(\ket{B_q},\ket{\bar B_q})$ is governed by two
hermitian $2\times 2$ matrices, the mass matrix $M^q$ and the decay
matrix $\Gamma^q$. By diagonalizing $M^q-i\Gamma^q/2$ one finds the mass
eigenstates $\ket{B_L^q}$ and $\ket{B_H^q}$ expressed in terms of the flavour
eigenstates $\ket{B_q}$, $\ket{\bar B_q}$. The mass eigenstates differ in
their masses $M_{H,L}^q$ and decay widths $\Gamma_{H,L}^q$ with ``L'' and
``H'' denoting ``light'' and ``heavy''. There are three observables, 
the mass and width
differences $\dm_q=M_H^q-M_L^q$ and $\dg_s=\Gamma_L^q-\Gamma_H^q$
as well as the CP asymmetry in flavor-specific decays, $a_{\rm fs}^q$.
Experimentally, $\dm_q$ is read off from the \bbq\ oscillation
frequency, $\dg_q$ is found by measuring lifetimes in 
different decay modes, and  $a_{\rm fs}^q$ is usually measured through
the time-dependent CP asymmetry in semileptonic $B_q$ decays. These
observables are related to the off-diagonal elements of $M^q$ and
$\Gamma^q$ as follows:
\begin{eqnarray}
\!\!\!  \dm_q  \simeq    2 |M_{12}^q|\,,\; 
  \frac{\Delta\Gamma_q}{\Delta M_q} =
  - \mbox{Re}\frac{\Gamma_{12}^q}{M_{12}^q} \,,\;
   a_{\rm fs}^q = \mbox{Im} \frac{\Gamma_{12}^q}{M_{12}^q}\,,\;
  \label{eq:dgdm}
\end{eqnarray}
with $|\Delta \Gamma_q| \simeq 2 |\Gamma_{12}^q|$.
$M_{12}^q$ is sensitive to new physics mediated by particles with masses well
beyond $100\tev$. On the contrary, $\Gamma_{12}^q$, probes effects of light
new particles with feeble couplings to quarks (see e.g.\
Refs.~\cite{Elor:2018twp,Alonso-Alvarez:2021qfd}).  While this is one
motivation for a more precise SM prediction of $\Gamma_{12}^q$, a better
knowledge of $\Gamma_{12}^q$ will also help to reveal new physics in
$M_{12}^q$: Inclusive and exclusive semileptonic $B$ decays give 
different values for the element $|V_{cb}|$ of the
Cabibbo-Kobayashi-Maskawa (CKM) matrix and this contoversy inflicts an
${\cal O}(15\%)$ uncertainty onto the overall CKM factor $(V_{tb}V_{tq}^*)^2$
of $M_{12}^q$.  This uncertainty drops out from the ratio $\dg_q/\dm_q$ in
\eq{eq:dgdm} and also the $4\%$ error from the hadronic matrix element in
$M_{12}^q$ largely cancels.  The measurements of LHCb~\cite{LHCb:2019nin},
CMS~\cite{CMS:2020efq}, ATLAS~\cite{ATLAS:2020lbz},
CDF~\cite{CDF:2012nqr}, and D\O~\cite{D0:2011ymu} combine to
\begin{eqnarray}
  \Delta\Gamma_s^{\rm exp}  &=&  (0.082 \pm 0.005)\; \mbox{ps}^{-1} 
                                \mbox{\cite{hfag}}
                                \label{eq:exp}\,,
\end{eqnarray}  
while $\dg_d^{\rm exp} $ is still consistent with zero. The precise
value in \eq{eq:exp} calls for a better SM prediction of $\dg_s$,
which is the topic of this Letter. We specify to $q=s$ from now on.

At one-loop level the SM predictions for $\Gamma_{12}^s$ is calculated from
the dispersive part of the $B_s \leftrightarrow \bar B_s$ amplitude.  One must
therefore only consider diagrams with light internal $u,c$ quarks; {\it i.e.}\
diagrams with one or two $t$ quarks only contribute to $M_{12}^s$. To properly
accomodate strong interaction effects associated with different energy scales
one employs two operator product expansions (OPE).  In the first step one
matches the SM to an effective theory with $|\Delta B|=1$ operators
\cite{Gilman:1979bc}, where $B$ is the beauty quantum number. The most
important operators, {\it i.e.} those with the largest coefficients, are the
current-current operators $Q_{1,2}$  describing tree-level $b$ decays.
The effective $|\Delta B|=1$ hamiltonian is known to next-to-leading (NLO)
\cite{Buras:1989xd,Buras:1990fn,Buras:1991jm} and next-to-next-to-leading
order (NNLO) \cite{Gorbahn:2004my,Gambino:2003zm,Gorbahn:2005sa} of Quantum
Chromodynamics (QCD).  The OPE employed in the second step of the calculation
is the Heavy Quark Expansion
(HQE)~\cite{Khoze:1983yp,Shifman:1984wx,Khoze:1986fa,Chay:1990da,Bigi:1991ir,Bigi:1992su,Bigi:1993fe,Blok:1993va,Manohar:1993qn}
(cf. also \cite{Lenz:2014jha} for a review), which expresses the
$B_s \leftrightarrow \bar B_s$ transition amplitude as a series in
$\lqcd/m_b$, where $\lqcd\sim 400\,$MeV is the fundamental scale of QCD and
$m_b$ is the $b$ quark mass. The HQE involves local $|\Delta B|=2$ operators; to
find the corresponding Wilson coefficients one must calculate the $\Delta B=2$
amplitude in both the $|\Delta B|=1$ and $|\Delta B|=2$ theories to the desired
order in $\alpha_s$.

The state of the art is as follows: QCD corrections to $\Gamma_{12}^s$ are
only known for the leading term of the $\lqcd/m_b$ expansion (``leading
power'').  These include NLO QCD corrections to the contributions with
current-current and chromomagnetic penguin operators
\cite{Beneke:1998sy,Ciuchini:2003ww,Beneke:2003az,Lenz:2006hd}, the
corresponding NNLO corrections (and NLO corrections involving four-quark
penguin operators) enhanced by the number $N_f$ of active quark 
flavours~\cite{Asatrian:2017qaz,Asatrian:2020zxa,Hovhannisyan:2022miy} as
well as NLO results with one current-current and one penguin operator
\cite{Gerlach:2021xtb} or two penguin operators \cite{Gerlach:2022wgb}. The
latter paper also presents two-loop results with one or two chromomagnetic
penguin operators which are part of the NNLO and N$^3$LO contributions.  (The
four-quark penguin operators $Q_{3-6}$ have Wilson coefficients which are much
smaller than those of $Q_{1,2}$ and the chromomagnetic penguin operator
contributes with a suppression factor of $\alpha_s$.)  The corrections of
Refs.~\cite{Asatrian:2017qaz} and
Refs.~\cite{Gerlach:2021xtb,Gerlach:2022wgb} have been calculated in an
expansion in $m_c/m_b$ to first and second order, respectively.  $\dg_s/\dm_s$
further involves a well-computed ratio of two hadronic matrix elements
\cite{Dowdall:2019bea,Kirk:2017juj,King:2021jsq}.  The contribution to
$\Gamma_{12}^s$ being sub-leading in $\lqcd/m_b$ is only known to LO of QCD
\cite{Beneke:1996gn} and the hadronic matrix elements still have large errors
\cite{Davies:2019gnp}.

Both the described perturbative contribution and the power-suppressed
term have theoretical uncertainties exceeding the experimental error in
\eq{eq:exp}. In this Letter we present  NNLO QCD corrections to the 
numerically dominant contribution with two current-current operators
and reduce the perturbative uncertainty of the leading-power term
to the level of the experimental error. 


\bigskip {\bf Calculation.}
To obtain $\Delta\Gamma_s / \Delta M_s$ we use the known two-loop QCD
corrections to $M_{12}^s$ from Ref.~\cite{Buras:1990fn}.
It is convenient to decompose $\Gamma_{12}^s$ according to the CKM structures
\begin{eqnarray}
  \Gamma_{12}^s &=& - (\lambda_c^s)^2\Gamma^{cc}_{12} 
                  - 2\lambda_c^s\lambda_u^s \Gamma_{12}^{uc} 
                  - (\lambda_u^s)^2\Gamma^{uu}_{12} 
                  \,,
                    \label{eq::Gam12}
\end{eqnarray}
where $\lambda^s_a = V_{as}^\ast V_{ab}$ with $a=u,c$.
$\Gamma_{12}^s$ is obtained with the help of a tower of effective theories. In a first step
we construct a theory where all degrees of freedom heavier than the bottom
quark mass $m_b$ are integrated out and the dynamical degrees of freedom are
given by the five lightest quarks and the gluons. We adopt the operator basis
of the $|\Delta B|=1$ theory
from Ref.~\cite{Chetyrkin:1997gb}.  The matching to the Standard Model happens
at the scale $\mu_0\approx 2 m_W \approx m_t(m_t)$. Afterwards, renormalization
group running is used to obtain the couplings of the effective operators at the
scale $\mu_1$ which is of the order $m_b$.

In a next step we perform a HQE
which allows us to write
$\Gamma_{12}^s$ as an expansion in $1/m_b$. At each order $\Gamma_{12}^s$ is
expressed as a sum of Wilson coefficients multiplying respective operator
matrix elements. The latter has to be computed using lattice gauge
theory~\cite{Dowdall:2019bea} or QCD sum rules~\cite{Kirk:2017juj,King:2021jsq}. To leading
order in the $1/m_b$ expansion we have
\begin{eqnarray}
  \Gamma_{12}^{ab} 
  &=& \frac{G_F^2m_b^2}{24\pi M_{B_s}} \left[ 
      H^{ab}(z)   \langle B_s|Q|\bar{B}_s \rangle
      + \widetilde{H}^{ab}_S(z)  \langle B_s|\widetilde{Q}_S|\bar{B}_s \rangle
      \right]
      \nonumber\\&&\mbox{}
      + {\cal O}(\Lambda_{\rm QCD}/m_b)\,,
      \label{eq::Gam^ab}
\end{eqnarray}
where $ab \in \{ cc,uc,uu \}$. $G_F$ is the Fermi constant and
$M_{B_s}$ is the mass of the $B_s$ meson.  The main purpose of this
Letter is the computation of the matching coefficients $H^{ab}$ and
$\widetilde{H}^{ab}_S$ to next-to-next-to-leading order (NNLO) in the strong
coupling constant $\alpha_s$.  They depend on $z=m_c^2/m_b^2$.  For the
$\Delta B=1$ theory one distinguishes current-current and penguin
operators. At leading and next-to-leading orders the current-current operators
provide about 90\% of the total contribution to
$\Gamma_{12}^{ab}$~\cite{Gerlach:2022wgb}. Thus, in this work we restrict
ourselves to the current-current contributions.

\begin{figure}[t]
  \begin{center}
    \begin{tabular}{cc}
      \includegraphics[width=0.23\textwidth]{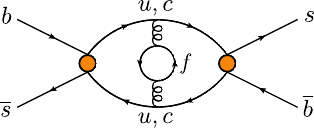}
      &
      \includegraphics[width=0.23\textwidth]{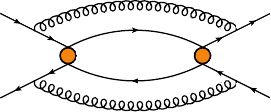}
      \\
      (a) & (b)
            \\
      \includegraphics[width=0.23\textwidth]{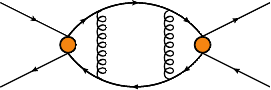}
      &
      \includegraphics[width=0.23\textwidth]{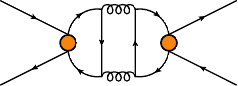}
      \\
      (c) & (d)
    \end{tabular}
    \caption{\label{fig::FD_DB1}Representative Feynman diagrams in the $\Delta B=1$ theory
      with $f=u,d,s,c,b$. Solid and curly lines represent quarks and
      gluons, respectively.  The (orange) blob indicates an operator
      insertion.}
  \end{center}
\end{figure}

For the calculation of the NNLO corrections one has to overcome several
challenges.  First, it is necessary to perform a three-loop calculation of the
amplitude $b\bar{s}\to \bar{b}s$ in the $\Delta B=1$ theory. Sample Feynman
diagrams are shown in Fig.~\ref{fig::FD_DB1}. In total about \mbox{20,000} three-loop
diagrams have to be considered which requires an automated setup for the
computation. In our case the combination of {\tt
  qgraf}~\cite{Nogueira:1991ex}, {\tt tapir}~\cite{Gerlach:2022qnc} and {\tt
  q2e/exp}~\cite{Harlander:1998cmq,Seidensticker:1999bb} turned out to be useful.
For the leading term in the HQE we are allowed to set the momentum of the
strange quark to zero. Furthermore, we expand in the charm quark mass up to
second order,\footnote{Up to this order a naive Taylor expansion of the
  amplitude is possible except for the fermionic corrections with a closed
  charm quark loop. These contributions are taken over from
  Ref.~\cite{Asatrian:2017qaz,Asatrian:2020zxa}.}  which reduces the integrals
to on-shell two-point functions with external momentum $q^2=m_b^2$.  The
propagators inside the loop diagrams are either massless or carry the mass
$m_b$. We use {\tt FIRE}~\cite{Smirnov:2019qkx} combined with {\tt
  LiteRed}~\cite{Lee:2012cn,Lee:2013mka} to reduce all occurring integrals to
23 genuine three-loop master integrals.  For the latter analytic results have been
obtained with the help of {\tt
  FeynCalc}~\cite{Mertig:1990an,Shtabovenko:2016sxi,Shtabovenko:2020gxv,Shtabovenko:2021hjx}, {\tt
  HyperInt}~\cite{Panzer:2015ida}, {\tt PolyLogTools}~\cite{Duhr:2019tlz} and
{\tt HyperlogProcedures}~\cite{schnetz}.

\begin{figure}[t]
  \begin{center}
    \begin{tabular}{ccc}
      \includegraphics[width=0.14\textwidth]{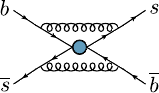}
      &
      \includegraphics[width=0.14\textwidth]{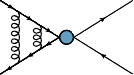}
      &
      \includegraphics[width=0.14\textwidth]{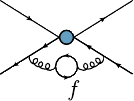}
      \\
      (a) & (b) & (c)
    \end{tabular}
    \caption{\label{fig::FD_DB2}Representative Feynman diagrams in the $\Delta B=2$
      theory. Solid and curly lines represent quarks and gluons,
      respectively. The (blue) blob indicates an operator insertion.}
  \end{center}
\end{figure}

On the $\Delta B=2$ side a two-loop calculation is necessary; sample Feynman
diagrams are shown in Fig.~\ref{fig::FD_DB2}. From the technical point of view
the calculation is significantly simpler. However, in the practical
calculation one has to consider three physical and 17 evanescent operators,
cf. Ref.~\cite{Gerlach:2022wgb}). It is necessary to compute the
corresponding renormalization constants for the operator mixing 
up to two-loop order.

The calculation of the $\Delta B=2$ matrix elements entails a field
theoretical subtlety.  In fact, in four dimensions there are only two physical
operators whereas for the calculation in $d$ dimensions three have to be taken
into account. For our calculation it is convenient to choose
$Q, \widetilde{Q}_S$ and $R_0$ where ($i$ and $j$ are colour indices)
\begin{eqnarray}
  Q &=& \bar{s}_i \gamma^\mu \,(1-\gamma^5)\, b_i \; \bar{s}_j \gamma_\mu
        \,(1-\gamma^5)\, b_j\,, \nonumber\\
  \widetilde{Q}_S &=& \bar{s}_i \,(1+\gamma^5)\, b_j\; \bar{s}_j \,(1+\gamma^5)\,
                  b_i\,,
          \label{eq::opDB2}
\end{eqnarray}
and
\begin{eqnarray}
  R_0 &=& Q_S + {\alpha_1} \widetilde Q_S + \frac{1}{2} {\alpha_2} Q\,,
          \label{eq::R0}
\end{eqnarray}  
with
\begin{eqnarray}
  Q_S &=& \bar{s}_i \,(1+\gamma^5)\, b_i \;\bar{s}_j \,(1+\gamma^5)\,
         b_j\,.
\end{eqnarray}
Note that at lowest order in $\alpha_s$ we have $\alpha_1=\alpha_2=1$ and the
matrix element of $R_0$ is $1/m_b$ suppressed in four dimensions. At higher
orders the quantities $\alpha_1$ and $\alpha_2$ are chosen such, that the
$1/m_b$-suppression is maintained. The one-loop corrections are known since
more than twenty years~\cite{Beneke:1998sy} and the fermionic
two-loop terms are available from Ref.~\cite{Asatrian:2017qaz}.  For the NNLO 
calculation performed in this Letter the $\alpha_s^2$ corrections to
$\alpha_1$ and $\alpha_2$ are needed.

The $1/m_b$-suppression of $R_0$ beyond tree-level is manifest only if
one is able to distinguish between ultraviolet (UV) and infrared (IR) divergences, e.g., by
regularizing the latter using a gluon mass $m_g$. Otherwise, $R_0$
develops an unphysical evanescent piece $E_{R_0}$ that scales as $m_b^0$
\cite{Gerlach:2022wgb} and hence must be included into the definition of
$R_0$ to obtain correct matching coefficients. One cannot isolate
  $E_{R_0}$ from $R_0$ at the operator level, but one can distinguish
evanescent and physical pieces in the matrix elements:
We use $R_0$ from Eq.~(\ref{eq::R0})
\emph{including} the finite UV renormalization encoded in
$\alpha_1$ and $\alpha_2$ in our matching calculation. To this end
we have first calculated the
linear combination of the renormalized
two-loop matrix elements $\langle Q \rangle^{(2)}$,
$\langle Q_S \rangle^{(2)}$ and $\langle \tilde{Q}_S \rangle^{(2)}$ as
given in Eq.~(\ref{eq::R0}).  After introducing a gluon mass along the
lines of Ref.~\cite{Chetyrkin:1997fm} and using Feynman gauge we observe
that each of the individual matrix elements becomes manifestly finite
upon UV renormalization.  $\alpha_{1}$ and $\alpha_{2}$ to order $\alpha_s^2$ are extracted
from the requirement that the linear combination must vanish in the
limit $m_b\to\infty$.

The matching between the $|\Delta B|=1$ and $|\Delta B|=2$ effective theories
is conceptually simple in case IR divergences are not regularized
dimensionally. In this case the UV renormalization renders amplitudes of both
theories manifestly finite, allowing us to take the limit $d \to 4$, where all
matrix elements of evanescent operators vanish. However, for technical reasons
we prefer to use $\epsilon = \epsilon_{\textrm{UV}} = \epsilon_{\textrm{IR}}$,
which simplifies the evaluation of the amplitudes but complicates the
matching. Following \cite{Ciuchini:2001vx} we need to extend the leading order
(LO) matching to $\mathcal{O}(\epsilon^2)$ and the NLO
matching to $\mathcal{O}(\epsilon)$ in order the determine the NNLO matching
coefficients.  Furthermore, we need to determine the matching coefficients of
both physical and evanescent operators: Since the UV-renormalized amplitudes
still contain IR poles, we must keep all matrix elements of evanescent operators
until the very end.  A powerful cross check of this procedure is the explicit
cancellation of the remaining IR $\epsilon$ poles and of the QCD gauge
parameter $\xi$ in the matching.


\bigskip {\bf Results.}  For our numerical analysis we use the input
values listed in Tab.~\ref{tab::input} and the $|\Delta B|=1$ Wilson
coefficients from
Refs.~\cite{Gorbahn:2004my,Gambino:2003zm,Gorbahn:2005sa} and calculate
the running and decoupling of quark masses and $\alpha_s$ with {\tt
  RunDec}~\cite{Herren:2017osy}.

\begin{table}[t]
  \begin{center}
    \renewcommand\arraystretch{1}
    \begin{tabular}{rclc}
      \hline 
      $\alpha_s(M_Z)$ &=& $0.1179 \pm 0.001$ & \cite{ParticleDataGroup:2020ssz}
      \\
      $m_c(3~\mbox{GeV})$ &=& $0.993\pm 0.008$~GeV & \cite{Chetyrkin:2017lif}
      \\
      $m_b(m_b)$ &=& $4.163\pm 0.016$~GeV & \cite{Chetyrkin:2017lif} 
      \\
      $m_t^{\rm pole}$ &=& $172.9\pm 0.4$~\mbox{GeV} & \cite{ParticleDataGroup:2020ssz} 
      \\
      $M_{B_s}$ &=& $5366.88$~\mbox{MeV} & \cite{ParticleDataGroup:2020ssz} 
      \\
      $B_{B_s}$ &=& $0.813\pm0.034$ & \cite{Dowdall:2019bea} 
      \\
      $\tilde{B}^\prime_{S,B_s}$ &=& $1.31\pm0.09$ & \cite{Dowdall:2019bea} 
      \\
      $f_{B_s}$ &=& $0.2307\pm0.0013$~GeV & \cite{Bazavov:2017lyh} 
      \\
      \hline 
    \end{tabular}
  \end{center}
  \caption{\label{tab::input}Input parameters for the numerical analysis. The
    matrix elements of $Q$ and $\widetilde Q_S$ are parametrized in terms of
    $f_{B_s}$, $B_{B_s}$, and $\tilde{B}^\prime_{S,B_s}$. The values of the
    quark masses imply $\bar z = 0.04956$, $m_b^{\rm pole}=4.75\gev$, and
    $m_b^{\rm PS}=4.479\gev$ (for a factorization scale $\mu_f=2$~GeV) at
    NNLO. Numerical results for the matrix elements of the $1/m_b$ suppressed
    corrections can be found in Ref.~\cite{Davies:2019gnp}.}
\end{table}

In the following we present the NNLO predictions in three different
renormalization schemes for the overall factor $m_b^2$
(cf. Eq.~(\ref{eq::Gam^ab})) whereas the quantity $z$ and the strong coupling
constant are defined in the ${\overline{\rm MS}}$ scheme. The overall factor
$m_b^2$ is defined in the ${\overline{\rm MS}}$ scheme, as a pole mass, or as
a potential-subtracted (PS) mass \cite{Beneke:1998rk}.  The latter is an
example of a so-called threshold mass, with similar properties as the pole
mass, but is nevertheless of short-distance nature. $H^{ab}(z)$ and
$\widetilde{H}^{ab}_S(z)$ are adapted accordingly, so that the scheme
dependence of $\Gamma_{12}^s$ is ${\cal O}(\alpha_s^3)$. Several
renormalization and matching scales enter the prediction for the width
difference. We choose $\mu_0=165$~GeV for the matching scale between the SM
and the $|\Delta B|=1$ theory.  Varying $\mu_0$ barely affects
$\Delta\Gamma_s/\Delta M_s$ and we can keep it fixed.  In our numerical
analysis we identify the matching scale $\mu_1$ and $\mu_b$ and $\mu_c$, the
renormalization scales at which $\overline{m}_{b}$ and $\overline{m}_{c}$ are
defined.  We simultaneously vary $\mu_1=\mu_b=\mu_c$ between $2.1$~GeV and
$8.4$~GeV with a central scale $\mu_1=4.2$~GeV. That is, $z$ enters the
coefficients as $\bar z= (m_c(\mu_1)/m_b(\mu_1))^2$.  The $|\Delta B|=2$
operators are defined at the scale $\mu_2$ which has to be kept fixed, because
the $\mu_2$ dependence only cancels in the products of $H^{ab}(z)$ and
$\widetilde{H}^{ab}_S(z)$ with their respective matrix elements.  In our
analysis we set $\mu_2=4.75$~GeV which is the bottom quark pole mass
$m_b^{\rm pole}$ obtained from $m_b(m_b)$ with two-loop accuracy.  The terms
of order $\lqcd/m_b$ in $\Gamma_{12}^s$ are only known to LO, so that the
$\mu_1$-dependence of these terms is non-negligible.

We now discuss the results for $\Delta\Gamma_s / \Delta M_s$.
In our three schemes we have
\begin{align}
  \frac{\Delta \Gamma_s}{\Delta M_s}
  ~=&~ \left(
      {3.79^{+0.53}_{-0.58}}_{\textrm{scale}}
      {{}^{+0.09}_{-0.19}}_{\textrm{scale, $1/m_b$}}
      \pm 0.11_{B\tilde{B}_S} \rt. \nonumber\\
  &~\lt. \; \pm 0.78_{1/m_b} \pm 0.05_{\textrm{input}}\right) \times
    10^{-3}\ (\textrm{pole})\,, \nonumber\\[1mm]
      \frac{\Delta \Gamma_s}{\Delta M_s} 
  ~=&~ \left(
      {4.33^{+0.23}_{ -0.44}}_{\textrm{scale}}
      {{}^{+0.09}_{ -0.19}}_{\textrm{scale, $1/m_b$}}
      \pm  0.12_{B\tilde{B}_S} \rt. \nonumber\\
  &~\lt. \;     \pm 0.78_{1/m_b} \pm 0.05_{\textrm{input}}\right) \times 10^{-3}\ (\overline{\textrm{MS}})\,,
   \nonumber\\[1mm]
  \frac{\Delta \Gamma_s}{\Delta M_s} ~=&~ 
 \left( { 4.20^{+0.36}_{ -0.39}}_{\textrm{scale}} 
      {{}^{ +0.09}_{ -0.19}}_{\textrm{scale, $1/m_b$}} \pm
          0.12_{B\tilde{B}_S}  \rt. \nonumber\\
  &~\lt. \;\pm 0.78_{1/m_b}
          \pm 0.05_{\textrm{input}}\right) \times 10^{-3}\ (\textrm{PS})\,,
          \label{eq::dGdM}
\end{align}
where the subscripts indicate the source of the various uncertainties.  The
dominant uncertainty comes from the matrix elements of the power-suppressed
corrections (``$1/m_b$'')~\cite{Davies:2019gnp,Dowdall:2019bea}) followed by
the renormalization scale uncertainty from the variation of $\mu_1$ in the
leading-power term (``scale'').  The uncertainties from the leading-power bag
parameters (``$B\tilde{B}_S$'') and from the scale variation in the $1/m_b$
piece (``scale, $1/m_b$'') are much smaller 
and the variation of the remaining input parameters (``input'') is of minor
relevance.

\begin{figure}[t]
  \begin{center}
    \includegraphics[width=0.45\textwidth]{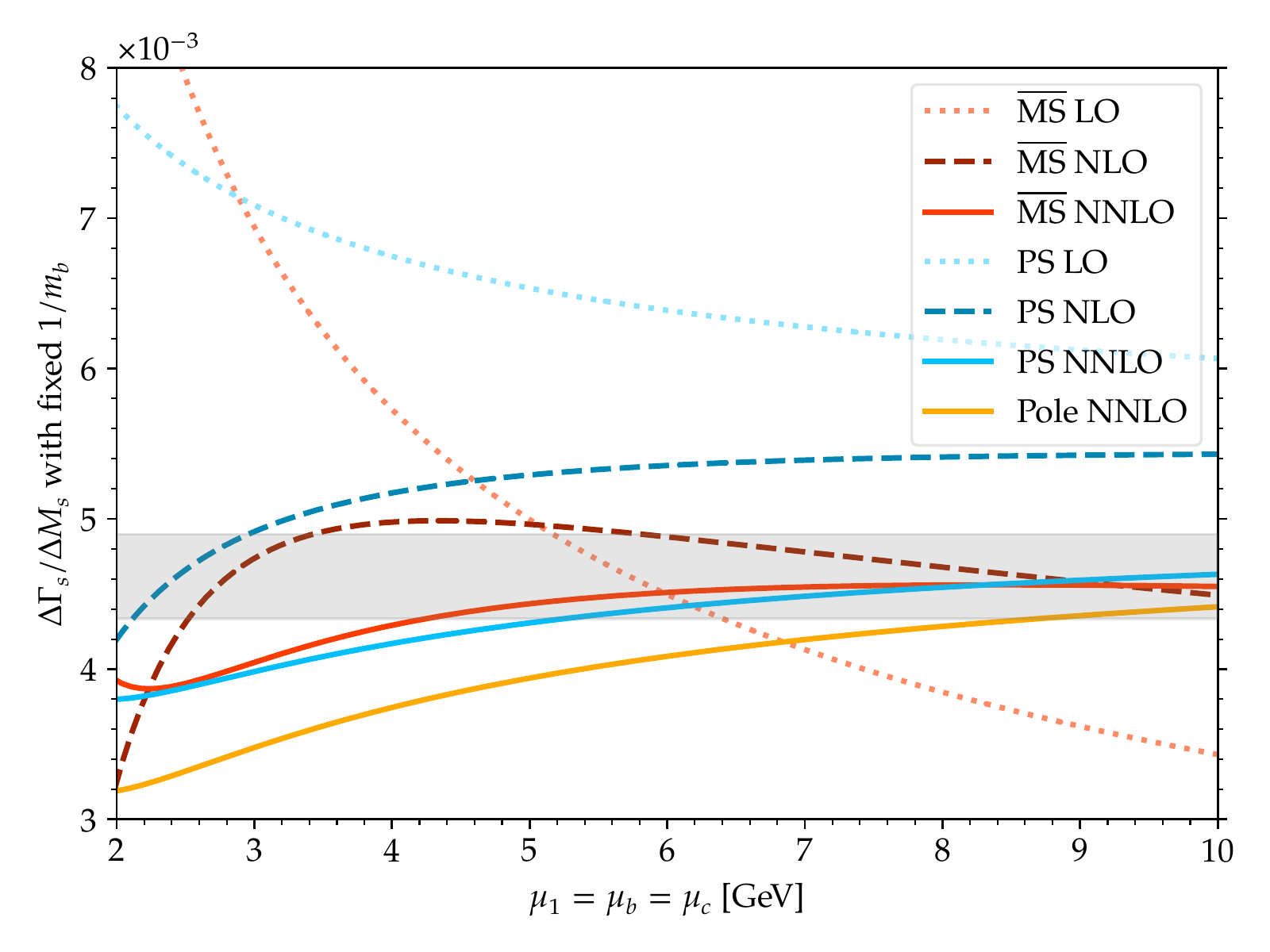}
    \caption{\label{fig::mu1}Renormalization scale dependence at LO, NLO and
      NNLO for the $\overline{\rm MS}$ and PS scheme. The scale
      in the power-suppressed terms is kept fixed. 
      The gray band represents the experimental result.}
  \end{center}
\end{figure}

In Fig.~\ref{fig::mu1} we show the dependence of $\dg_s/\dm_s$ on the
simultaneously varied renormalization scales $\mu_1=\mu_b=\mu_c$ for the
$\overline{\rm MS}$ and PS schemes. The small contributions involving
four-quark penguin operators are only included at NLO in both the NLO and NNLO
curves. Dotted, dashed, and solid curves correspond to the LO, NLO, and NNLO
results, respectively. In both schemes one observes a clear stabilization of
the $\mu_1$ dependence after including higher orders. Furthermore, we observe
that the NNLO predictions (solid lines) in both schemes are close together
which demonstrates the expected reduction of the scheme dependence.  In the
$\overline{\rm MS}$ scheme we observe that the LO and NLO curves intersect
close to the central scale. As a consequence the NLO corrections are
relatively small and the NNLO contributions are of comparable size. Close to
9~GeV the NNLO contribution is zero and the NLO corrections amount to about
$+21$\%. At the same time the NNLO predictions for $\mu_1=4.2$~GeV and
$\mu_1=9$~GeV differ only by $+5$\% and $+9$\% in the $\ov{\rm MS}$ and PS
schemes, respectively. Note that in the $\overline{\rm MS}$ scheme the scale
dependence of the leading-power term drops from {${}^{+0}_{-29}$\%} at NLO
to {${}^{+5}_{-10}$\%} at NNLO and is now of the same order of
magnitude as the $\pm 6\%$ experimental error in~\eq{eq:exp}. In the PS
scheme the scale uncertainty is of the same order of magnitude as in the
$\overline{\rm MS}$ scheme. Note that the scheme dependence inferred from
the $\ov{\rm MS}$ and PS central values in \eq{eq::dGdM} is only 3\%.
\eq{eq::dGdM} clearly shows that one needs better results for the $1/m_b$
matrix elements. A meaningful lattice-continuum matching calls for NLO
corrections to the power-suppressed terms, which will further reduce the
uncertainty labeled with ``scale, $1/m_b$''.

For the pole scheme we only show the NNLO prediction in Fig.~\ref{fig::mu1}.
While we also see a relatively mild dependence on $\mu_1$, the corresponding
solid curve lies significantly below the predictions in the
$\overline{\rm MS}$ and PS schemes.  This feature can be traced back to the
large two-loop corrections in the relation between the $\overline{\rm MS}$ and
the pole bottom quark mass affecting NNLO contributions as much as the genuine
NNLO corrections, underpinning the well-known issues with quark pole masses
\cite{Bigi:1994em,Beneke:1994sw,Beneke:2021lkq}.  For this reason we recommend
to not use the pole scheme for the prediction of $\Delta\Gamma_s$.

The most precise prediction for $\Delta\Gamma_s$ is obtained from the results
in Eq.~(\ref{eq::dGdM}) combined with the experimental
result~\cite{LHCb:2021moh}
$\Delta M_s^{\rm exp} = 17.7656 \pm 0.0057~\mbox{ps}^{-1}$.  Upon adding the
various uncertainties in quadrature, symmetrizing the scale dependence and
averaging the results from the $\overline{\rm MS}$ and PS schemes we obtain
\begin{eqnarray}
  \Delta\Gamma_s &=& {(0.076 \pm 0.017)} ~\mbox{ps}^{-1}\,.
\end{eqnarray}
The comparison to Eq.~(\ref{eq:exp}) shows that the uncertainty is only about
a factor three bigger than from experiment and dominated by
the $1/m_b$ corrections.

With our NNLO result for $\Gamma_{12}^q$ we can also improve the
predictions for width difference in the $B_d-\bar{B}_d$ system and the
CP asymmetries $a_{\rm fs}^s$ and $a_{\rm fs}^d$ in \eq{eq:dgdm},
whose experimental results are still consistent with zero. We
postpone this to a future publication~\cite{Gerlach_in_prep}.

\begin{figure}[t]
  \begin{center}
    \includegraphics[width=0.45\textwidth]{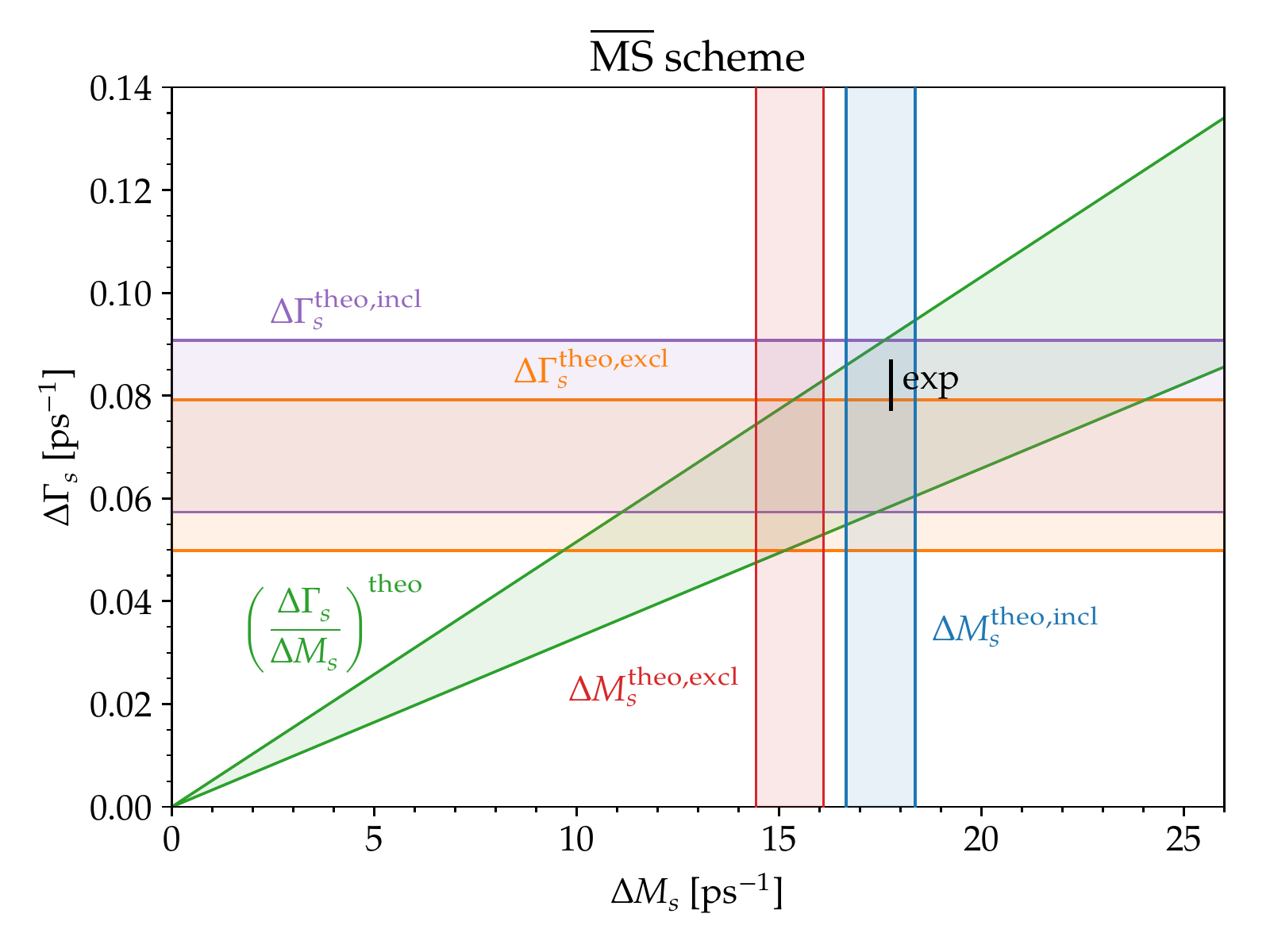}
    \caption{\label{fig::dGdM}$\Delta\Gamma_s$ versus $\Delta M_s$. The
      $|V_{cb}|$ controversy (red vs. blue vertical and orange vs. purple
      horizontal strips) prevents any
      conclusion on possible new physics in $\dm_s$. A combined analysis of
      $\dm_s$ and $\dg_s$ adds important information, because the SM
      prediction of $\dg_s/\dm_s$ (green wedge) is independent of $|V_{cb}|$.
    }
  \end{center}
\end{figure}

In Fig.~\ref{fig::dGdM} we confront our predictions for the ratio
$\Delta\Gamma_s/\Delta M_s$ in the $\overline{\rm MS}$ scheme
(green band) with
the individual predictions of $\Delta\Gamma_s$ and $\Delta M_s$.  The
latter are dominated by the uncertainty in the CKM matrix element
$V_{ts}$ which is obtained from $V_{cb}$ through CKM unitarity and
cancels in the ratio.
Fig.~\ref{fig::dGdM} illustrates this feature with
$|V_{cb}^{\rm incl}|=42.16(51)10^{-3}$ from~\cite{Bordone:2021oof} and
$|V_{cb}^{\rm excl}|=39.36(68)10^{-3}$ from~\cite{Aoki:2021kgd}.
The current experimental results for $\Delta\Gamma_s$ and $\Delta M_s$ are
indicated by the black bar. Once the prediction of $\dg_s/\dm_s$ is
improved further, it will be possible to test the SM without CKM
uncertainty, and with progress on $|V_{cb}|$ one will be able to
constrain new physics in $\dm_s$ and $\dg_s$ individually.


\bigskip {\bf Conclusions.} The SM prediction of $\dg_s/\dm_s$ based
  on the long-standing NLO calculation has two sources of uncertainty
  which exceed the experimental error: the hadronic matrix elements of
  the power-suppressed operators and the perturbative coefficients, as
  inferred from the scale and scheme dependences of the calculated
  result. With the NNLO calculation presented here we have brought the
  latter uncertainty to the level of the accuracy of the experimental
  result. For this we had to calculate \mbox{20,000} three-loop diagrams and to
  solve subtle problems related to the interplay of infrared divergences
  and evanescent oprators. We have pointed out that $\dg_s$ adds
  information to the usual study of $\dm_s$, because both quantities
  probe different new-physics scenarios and $|V_{cb}|$ drops out in the
  ratio $\dg_s/\dm_s$.


\smallskip

{\bf Acknowledgements.}  We thank Artyom Hovhannisyan and Matthew Wingate for
useful discussions and we are grateful to Erik Panzer and Oliver Schnetz for
helpful advice regarding the calculation of the master integrals and the
simplification of the obtained results with {\tt HyperInt} and {\tt
  HyperLogProcedures}.  VS thanks David Broadhurst for enlightening
discussions on iterated integrals. This research was supported by the
Deutsche Forschungsgemeinschaft (DFG, German Research Foundation) under grant
396021762 --- TRR 257 ``Particle Physics Phenomenology after the Higgs
Discovery''. The Feynman diagrams were drawn with the help of
Axodraw~\cite{Vermaseren:1994je} and JaxoDraw~\cite{Binosi:2003yf}.

\bibliography{inspire_extra}

\begin{thebibliography}{70}%
\makeatletter
\providecommand \@ifxundefined [1]{%
 \@ifx{#1\undefined}
}%
\providecommand \@ifnum [1]{%
 \ifnum #1\expandafter \@firstoftwo
 \else \expandafter \@secondoftwo
 \fi
}%
\providecommand \@ifx [1]{%
 \ifx #1\expandafter \@firstoftwo
 \else \expandafter \@secondoftwo
 \fi
}%
\providecommand \natexlab [1]{#1}%
\providecommand \enquote  [1]{``#1''}%
\providecommand \bibnamefont  [1]{#1}%
\providecommand \bibfnamefont [1]{#1}%
\providecommand \citenamefont [1]{#1}%
\providecommand \href@noop [0]{\@secondoftwo}%
\providecommand \href [0]{\begingroup \@sanitize@url \@href}%
\providecommand \@href[1]{\@@startlink{#1}\@@href}%
\providecommand \@@href[1]{\endgroup#1\@@endlink}%
\providecommand \@sanitize@url [0]{\catcode `\\12\catcode `\$12\catcode
  `\&12\catcode `\#12\catcode `\^12\catcode `\_12\catcode `\%12\relax}%
\providecommand \@@startlink[1]{}%
\providecommand \@@endlink[0]{}%
\providecommand \url  [0]{\begingroup\@sanitize@url \@url }%
\providecommand \@url [1]{\endgroup\@href {#1}{\urlprefix }}%
\providecommand \urlprefix  [0]{URL }%
\providecommand \Eprint [0]{\href }%
\providecommand \doibase [0]{http://dx.doi.org/}%
\providecommand \selectlanguage [0]{\@gobble}%
\providecommand \bibinfo  [0]{\@secondoftwo}%
\providecommand \bibfield  [0]{\@secondoftwo}%
\providecommand \translation [1]{[#1]}%
\providecommand \BibitemOpen [0]{}%
\providecommand \bibitemStop [0]{}%
\providecommand \bibitemNoStop [0]{.\EOS\space}%
\providecommand \EOS [0]{\spacefactor3000\relax}%
\providecommand \BibitemShut  [1]{\csname bibitem#1\endcsname}%
\let\auto@bib@innerbib\@empty
\bibitem [{\citenamefont {Elor}\ \emph {et~al.}(2019)\citenamefont {Elor},
  \citenamefont {Escudero},\ and\ \citenamefont {Nelson}}]{Elor:2018twp}%
  \BibitemOpen
  \bibfield  {author} {\bibinfo {author} {\bibfnamefont {Gilly}\ \bibnamefont
  {Elor}}, \bibinfo {author} {\bibfnamefont {Miguel}\ \bibnamefont {Escudero}},
  \ and\ \bibinfo {author} {\bibfnamefont {Ann}\ \bibnamefont {Nelson}},\
  }\bibfield  {title} {\enquote {\bibinfo {title} {{Baryogenesis and Dark
  Matter from $B$ Mesons}},}\ }\href {\doibase 10.1103/PhysRevD.99.035031}
  {\bibfield  {journal} {\bibinfo  {journal} {Phys. Rev. D}\ }\textbf {\bibinfo
  {volume} {99}},\ \bibinfo {pages} {035031} (\bibinfo {year} {2019})},\
  \Eprint {http://arxiv.org/abs/1810.00880} {arXiv:1810.00880 [hep-ph]}
  \BibitemShut {NoStop}%
\bibitem [{\citenamefont {Alonso-\'Alvarez}\ \emph {et~al.}(2021)\citenamefont
  {Alonso-\'Alvarez}, \citenamefont {Elor},\ and\ \citenamefont
  {Escudero}}]{Alonso-Alvarez:2021qfd}%
  \BibitemOpen
  \bibfield  {author} {\bibinfo {author} {\bibfnamefont {Gonzalo}\ \bibnamefont
  {Alonso-\'Alvarez}}, \bibinfo {author} {\bibfnamefont {Gilly}\ \bibnamefont
  {Elor}}, \ and\ \bibinfo {author} {\bibfnamefont {Miguel}\ \bibnamefont
  {Escudero}},\ }\bibfield  {title} {\enquote {\bibinfo {title} {{Collider
  signals of baryogenesis and dark matter from B mesons: A roadmap to
  discovery}},}\ }\href {\doibase 10.1103/PhysRevD.104.035028} {\bibfield
  {journal} {\bibinfo  {journal} {Phys. Rev. D}\ }\textbf {\bibinfo {volume}
  {104}},\ \bibinfo {pages} {035028} (\bibinfo {year} {2021})},\ \Eprint
  {http://arxiv.org/abs/2101.02706} {arXiv:2101.02706 [hep-ph]} \BibitemShut
  {NoStop}%
\bibitem [{\citenamefont {Aaij}\ \emph {et~al.}(2019)\citenamefont {Aaij} \emph
  {et~al.}}]{LHCb:2019nin}%
  \BibitemOpen
  \bibfield  {author} {\bibinfo {author} {\bibfnamefont {Roel}\ \bibnamefont
  {Aaij}} \emph {et~al.} (\bibinfo {collaboration} {LHCb}),\ }\bibfield
  {title} {\enquote {\bibinfo {title} {{Updated measurement of time-dependent
  {\textbackslash{}it CP}-violating observables in $B^{0}_{s}\to J/\psi K^+
  K^-$ decays}},}\ }\href {\doibase 10.1140/epjc/s10052-019-7159-8} {\bibfield
  {journal} {\bibinfo  {journal} {Eur. Phys. J. C}\ }\textbf {\bibinfo {volume}
  {79}},\ \bibinfo {pages} {706} (\bibinfo {year} {2019})},\ \bibinfo {note}
  {[Erratum: Eur.Phys.J.C 80, 601 (2020)]},\ \Eprint
  {http://arxiv.org/abs/1906.08356} {arXiv:1906.08356 [hep-ex]} \BibitemShut
  {NoStop}%
\bibitem [{\citenamefont {Sirunyan}\ \emph {et~al.}(2021)\citenamefont
  {Sirunyan} \emph {et~al.}}]{CMS:2020efq}%
  \BibitemOpen
  \bibfield  {author} {\bibinfo {author} {\bibfnamefont {Albert~M}\
  \bibnamefont {Sirunyan}} \emph {et~al.} (\bibinfo {collaboration} {CMS}),\
  }\bibfield  {title} {\enquote {\bibinfo {title} {{Measurement of the
  $CP$-violating phase $\phi_\mathrm{s}$ in the B$^0_\mathrm{s}\to$ J$/\psi\,
  \phi$(1020) $\to \mu^+\mu^-$K$^+$K$^-$ channel in proton-proton collisions at
  $\sqrt{s} =$ 13 TeV}},}\ }\href {\doibase 10.1016/j.physletb.2021.136188}
  {\bibfield  {journal} {\bibinfo  {journal} {Phys. Lett. B}\ }\textbf
  {\bibinfo {volume} {816}},\ \bibinfo {pages} {136188} (\bibinfo {year}
  {2021})},\ \Eprint {http://arxiv.org/abs/2007.02434} {arXiv:2007.02434
  [hep-ex]} \BibitemShut {NoStop}%
\bibitem [{\citenamefont {Aad}\ \emph {et~al.}(2021)\citenamefont {Aad} \emph
  {et~al.}}]{ATLAS:2020lbz}%
  \BibitemOpen
  \bibfield  {author} {\bibinfo {author} {\bibfnamefont {Georges}\ \bibnamefont
  {Aad}} \emph {et~al.} (\bibinfo {collaboration} {ATLAS}),\ }\bibfield
  {title} {\enquote {\bibinfo {title} {{Measurement of the $CP$-violating phase
  $\phi_s$ in $B^0_s \to J/\psi\phi$ decays in ATLAS at 13 TeV}},}\ }\href
  {\doibase 10.1140/epjc/s10052-021-09011-0} {\bibfield  {journal} {\bibinfo
  {journal} {Eur. Phys. J. C}\ }\textbf {\bibinfo {volume} {81}},\ \bibinfo
  {pages} {342} (\bibinfo {year} {2021})},\ \Eprint
  {http://arxiv.org/abs/2001.07115} {arXiv:2001.07115 [hep-ex]} \BibitemShut
  {NoStop}%
\bibitem [{\citenamefont {Aaltonen}\ \emph {et~al.}(2012)\citenamefont
  {Aaltonen} \emph {et~al.}}]{CDF:2012nqr}%
  \BibitemOpen
  \bibfield  {author} {\bibinfo {author} {\bibfnamefont {T.}~\bibnamefont
  {Aaltonen}} \emph {et~al.} (\bibinfo {collaboration} {CDF}),\ }\bibfield
  {title} {\enquote {\bibinfo {title} {{Measurement of the Bottom-Strange Meson
  Mixing Phase in the Full CDF Data Set}},}\ }\href {\doibase
  10.1103/PhysRevLett.109.171802} {\bibfield  {journal} {\bibinfo  {journal}
  {Phys. Rev. Lett.}\ }\textbf {\bibinfo {volume} {109}},\ \bibinfo {pages}
  {171802} (\bibinfo {year} {2012})},\ \Eprint {http://arxiv.org/abs/1208.2967}
  {arXiv:1208.2967 [hep-ex]} \BibitemShut {NoStop}%
\bibitem [{\citenamefont {Abazov}\ \emph {et~al.}(2012)\citenamefont {Abazov}
  \emph {et~al.}}]{D0:2011ymu}%
  \BibitemOpen
  \bibfield  {author} {\bibinfo {author} {\bibfnamefont {Victor~Mukhamedovich}\
  \bibnamefont {Abazov}} \emph {et~al.} (\bibinfo {collaboration} {D0}),\
  }\bibfield  {title} {\enquote {\bibinfo {title} {{Measurement of the
  CP-violating phase $\phi_s^{J/\psi \phi}$ using the flavor-tagged decay
  $B_s^0 \rightarrow J/\psi \phi$ in 8 fb$^{-1}$ of $p \bar p$ collisions}},}\
  }\href {\doibase 10.1103/PhysRevD.85.032006} {\bibfield  {journal} {\bibinfo
  {journal} {Phys. Rev. D}\ }\textbf {\bibinfo {volume} {85}},\ \bibinfo
  {pages} {032006} (\bibinfo {year} {2012})},\ \Eprint
  {http://arxiv.org/abs/1109.3166} {arXiv:1109.3166 [hep-ex]} \BibitemShut
  {NoStop}%
\bibitem [{\citenamefont {(HFLAV)}()}]{hfag}%
  \BibitemOpen
  \bibfield  {author} {\bibinfo {author} {\bibfnamefont {Heavy Flavor
  Averaging~Group}\ \bibnamefont {(HFLAV)}},\ }\bibfield  {title} {\enquote
  {\bibinfo {title} {online update at},}\ }\href@noop {} {\bibinfo  {journal}
  {\url{https://hflav-eos.web.cern.ch/hflav-eos/osc/PDG_2020/\# DMS}}\
  }\BibitemShut {NoStop}%
\bibitem [{\citenamefont {Gilman}\ and\ \citenamefont
  {Wise}(1979)}]{Gilman:1979bc}%
  \BibitemOpen
\bibfield  {journal} {  }\bibfield  {author} {\bibinfo {author} {\bibfnamefont
  {Frederick~J.}\ \bibnamefont {Gilman}}\ and\ \bibinfo {author} {\bibfnamefont
  {Mark~B.}\ \bibnamefont {Wise}},\ }\bibfield  {title} {\enquote {\bibinfo
  {title} {{Effective Hamiltonian for Delta s = 1 Weak Nonleptonic Decays in
  the Six Quark Model}},}\ }\href {\doibase 10.1103/PhysRevD.20.2392}
  {\bibfield  {journal} {\bibinfo  {journal} {Phys. Rev. D}\ }\textbf {\bibinfo
  {volume} {20}},\ \bibinfo {pages} {2392} (\bibinfo {year}
  {1979})}\BibitemShut {NoStop}%
\bibitem [{\citenamefont {Buras}\ and\ \citenamefont
  {Weisz}(1990)}]{Buras:1989xd}%
  \BibitemOpen
  \bibfield  {author} {\bibinfo {author} {\bibfnamefont {Andrzej~J.}\
  \bibnamefont {Buras}}\ and\ \bibinfo {author} {\bibfnamefont {Peter~H.}\
  \bibnamefont {Weisz}},\ }\bibfield  {title} {\enquote {\bibinfo {title} {{QCD
  Nonleading Corrections to Weak Decays in Dimensional Regularization and 't
  Hooft-Veltman Schemes}},}\ }\href {\doibase 10.1016/0550-3213(90)90223-Z}
  {\bibfield  {journal} {\bibinfo  {journal} {Nucl. Phys. B}\ }\textbf
  {\bibinfo {volume} {333}},\ \bibinfo {pages} {66--99} (\bibinfo {year}
  {1990})}\BibitemShut {NoStop}%
\bibitem [{\citenamefont {Buras}\ \emph {et~al.}(1990)\citenamefont {Buras},
  \citenamefont {Jamin},\ and\ \citenamefont {Weisz}}]{Buras:1990fn}%
  \BibitemOpen
  \bibfield  {author} {\bibinfo {author} {\bibfnamefont {Andrzej~J.}\
  \bibnamefont {Buras}}, \bibinfo {author} {\bibfnamefont {Matthias}\
  \bibnamefont {Jamin}}, \ and\ \bibinfo {author} {\bibfnamefont {Peter~H.}\
  \bibnamefont {Weisz}},\ }\bibfield  {title} {\enquote {\bibinfo {title}
  {{Leading and Next-to-leading {QCD} Corrections to $\epsilon$ Parameter and
  $B^0 - \bar{B}^0$ Mixing in the Presence of a Heavy Top Quark}},}\ }\href
  {\doibase 10.1016/0550-3213(90)90373-L} {\bibfield  {journal} {\bibinfo
  {journal} {Nucl. Phys. B}\ }\textbf {\bibinfo {volume} {347}},\ \bibinfo
  {pages} {491--536} (\bibinfo {year} {1990})}\BibitemShut {NoStop}%
\bibitem [{\citenamefont {Buras}\ \emph {et~al.}(1992)\citenamefont {Buras},
  \citenamefont {Jamin}, \citenamefont {Lautenbacher},\ and\ \citenamefont
  {Weisz}}]{Buras:1991jm}%
  \BibitemOpen
  \bibfield  {author} {\bibinfo {author} {\bibfnamefont {Andrzej~J.}\
  \bibnamefont {Buras}}, \bibinfo {author} {\bibfnamefont {Matthias}\
  \bibnamefont {Jamin}}, \bibinfo {author} {\bibfnamefont {M.~E.}\ \bibnamefont
  {Lautenbacher}}, \ and\ \bibinfo {author} {\bibfnamefont {Peter~H.}\
  \bibnamefont {Weisz}},\ }\bibfield  {title} {\enquote {\bibinfo {title}
  {{Effective Hamiltonians for $\Delta S = 1$ and $\Delta B = 1$ nonleptonic
  decays beyond the leading logarithmic approximation}},}\ }\href {\doibase
  10.1016/0550-3213(92)90345-C} {\bibfield  {journal} {\bibinfo  {journal}
  {Nucl. Phys. B}\ }\textbf {\bibinfo {volume} {370}},\ \bibinfo {pages}
  {69--104} (\bibinfo {year} {1992})},\ \bibinfo {note} {[Addendum: Nucl.Phys.B
  375, 501 (1992)]}\BibitemShut {NoStop}%
\bibitem [{\citenamefont {Gorbahn}\ and\ \citenamefont
  {Haisch}(2005)}]{Gorbahn:2004my}%
  \BibitemOpen
  \bibfield  {author} {\bibinfo {author} {\bibfnamefont {Martin}\ \bibnamefont
  {Gorbahn}}\ and\ \bibinfo {author} {\bibfnamefont {Ulrich}\ \bibnamefont
  {Haisch}},\ }\bibfield  {title} {\enquote {\bibinfo {title} {{Effective
  Hamiltonian for non-leptonic $|\Delta F| = 1$ decays at NNLO in QCD}},}\
  }\href {\doibase 10.1016/j.nuclphysb.2005.01.047} {\bibfield  {journal}
  {\bibinfo  {journal} {Nucl. Phys. B}\ }\textbf {\bibinfo {volume} {713}},\
  \bibinfo {pages} {291--332} (\bibinfo {year} {2005})},\ \Eprint
  {http://arxiv.org/abs/hep-ph/0411071} {arXiv:hep-ph/0411071} \BibitemShut
  {NoStop}%
\bibitem [{\citenamefont {Gambino}\ \emph {et~al.}(2003)\citenamefont
  {Gambino}, \citenamefont {Gorbahn},\ and\ \citenamefont
  {Haisch}}]{Gambino:2003zm}%
  \BibitemOpen
  \bibfield  {author} {\bibinfo {author} {\bibfnamefont {Paolo}\ \bibnamefont
  {Gambino}}, \bibinfo {author} {\bibfnamefont {Martin}\ \bibnamefont
  {Gorbahn}}, \ and\ \bibinfo {author} {\bibfnamefont {Ulrich}\ \bibnamefont
  {Haisch}},\ }\bibfield  {title} {\enquote {\bibinfo {title} {{Anomalous
  dimension matrix for radiative and rare semileptonic B decays up to three
  loops}},}\ }\href {\doibase 10.1016/j.nuclphysb.2003.09.024} {\bibfield
  {journal} {\bibinfo  {journal} {Nucl. Phys. B}\ }\textbf {\bibinfo {volume}
  {673}},\ \bibinfo {pages} {238--262} (\bibinfo {year} {2003})},\ \Eprint
  {http://arxiv.org/abs/hep-ph/0306079} {arXiv:hep-ph/0306079} \BibitemShut
  {NoStop}%
\bibitem [{\citenamefont {Gorbahn}\ \emph {et~al.}(2005)\citenamefont
  {Gorbahn}, \citenamefont {Haisch},\ and\ \citenamefont
  {Misiak}}]{Gorbahn:2005sa}%
  \BibitemOpen
  \bibfield  {author} {\bibinfo {author} {\bibfnamefont {Martin}\ \bibnamefont
  {Gorbahn}}, \bibinfo {author} {\bibfnamefont {Ulrich}\ \bibnamefont
  {Haisch}}, \ and\ \bibinfo {author} {\bibfnamefont {Mikolaj}\ \bibnamefont
  {Misiak}},\ }\bibfield  {title} {\enquote {\bibinfo {title} {{Three-loop
  mixing of dipole operators}},}\ }\href {\doibase
  10.1103/PhysRevLett.95.102004} {\bibfield  {journal} {\bibinfo  {journal}
  {Phys. Rev. Lett.}\ }\textbf {\bibinfo {volume} {95}},\ \bibinfo {pages}
  {102004} (\bibinfo {year} {2005})},\ \Eprint
  {http://arxiv.org/abs/hep-ph/0504194} {arXiv:hep-ph/0504194} \BibitemShut
  {NoStop}%
\bibitem [{\citenamefont {Khoze}\ and\ \citenamefont
  {Shifman}(1983)}]{Khoze:1983yp}%
  \BibitemOpen
  \bibfield  {author} {\bibinfo {author} {\bibfnamefont {Valery~A.}\
  \bibnamefont {Khoze}}\ and\ \bibinfo {author} {\bibfnamefont {Mikhail~A.}\
  \bibnamefont {Shifman}},\ }\bibfield  {title} {\enquote {\bibinfo {title}
  {{HEAVY QUARKS}},}\ }\href {\doibase 10.1070/PU1983v026n05ABEH004398}
  {\bibfield  {journal} {\bibinfo  {journal} {Sov. Phys. Usp.}\ }\textbf
  {\bibinfo {volume} {26}},\ \bibinfo {pages} {387} (\bibinfo {year}
  {1983})}\BibitemShut {NoStop}%
\bibitem [{\citenamefont {Shifman}\ and\ \citenamefont
  {Voloshin}(1985)}]{Shifman:1984wx}%
  \BibitemOpen
  \bibfield  {author} {\bibinfo {author} {\bibfnamefont {Mikhail~A.}\
  \bibnamefont {Shifman}}\ and\ \bibinfo {author} {\bibfnamefont {M.~B.}\
  \bibnamefont {Voloshin}},\ }\bibfield  {title} {\enquote {\bibinfo {title}
  {{Preasymptotic Effects in Inclusive Weak Decays of Charmed Particles}},}\
  }\href@noop {} {\bibfield  {journal} {\bibinfo  {journal} {Sov. J. Nucl.
  Phys.}\ }\textbf {\bibinfo {volume} {41}},\ \bibinfo {pages} {120} (\bibinfo
  {year} {1985})}\BibitemShut {NoStop}%
\bibitem [{\citenamefont {Khoze}\ \emph {et~al.}(1987)\citenamefont {Khoze},
  \citenamefont {Shifman}, \citenamefont {Uraltsev},\ and\ \citenamefont
  {Voloshin}}]{Khoze:1986fa}%
  \BibitemOpen
  \bibfield  {author} {\bibinfo {author} {\bibfnamefont {Valery~A.}\
  \bibnamefont {Khoze}}, \bibinfo {author} {\bibfnamefont {Mikhail~A.}\
  \bibnamefont {Shifman}}, \bibinfo {author} {\bibfnamefont {N.~G.}\
  \bibnamefont {Uraltsev}}, \ and\ \bibinfo {author} {\bibfnamefont {M.~B.}\
  \bibnamefont {Voloshin}},\ }\bibfield  {title} {\enquote {\bibinfo {title}
  {{On Inclusive Hadronic Widths of Beautiful Particles}},}\ }\href@noop {}
  {\bibfield  {journal} {\bibinfo  {journal} {Sov. J. Nucl. Phys.}\ }\textbf
  {\bibinfo {volume} {46}},\ \bibinfo {pages} {112} (\bibinfo {year}
  {1987})}\BibitemShut {NoStop}%
\bibitem [{\citenamefont {Chay}\ \emph {et~al.}(1990)\citenamefont {Chay},
  \citenamefont {Georgi},\ and\ \citenamefont {Grinstein}}]{Chay:1990da}%
  \BibitemOpen
  \bibfield  {author} {\bibinfo {author} {\bibfnamefont {Junegone}\
  \bibnamefont {Chay}}, \bibinfo {author} {\bibfnamefont {Howard}\ \bibnamefont
  {Georgi}}, \ and\ \bibinfo {author} {\bibfnamefont {Benjamin}\ \bibnamefont
  {Grinstein}},\ }\bibfield  {title} {\enquote {\bibinfo {title} {{Lepton
  energy distributions in heavy meson decays from QCD}},}\ }\href {\doibase
  10.1016/0370-2693(90)90916-T} {\bibfield  {journal} {\bibinfo  {journal}
  {Phys. Lett. B}\ }\textbf {\bibinfo {volume} {247}},\ \bibinfo {pages}
  {399--405} (\bibinfo {year} {1990})}\BibitemShut {NoStop}%
\bibitem [{\citenamefont {Bigi}\ and\ \citenamefont
  {Uraltsev}(1992)}]{Bigi:1991ir}%
  \BibitemOpen
  \bibfield  {author} {\bibinfo {author} {\bibfnamefont {Ikaros I.~Y.}\
  \bibnamefont {Bigi}}\ and\ \bibinfo {author} {\bibfnamefont {N.~G.}\
  \bibnamefont {Uraltsev}},\ }\bibfield  {title} {\enquote {\bibinfo {title}
  {{Gluonic enhancements in non-spectator beauty decays: An Inclusive mirage
  though an exclusive possibility}},}\ }\href {\doibase
  10.1016/0370-2693(92)90066-D} {\bibfield  {journal} {\bibinfo  {journal}
  {Phys. Lett. B}\ }\textbf {\bibinfo {volume} {280}},\ \bibinfo {pages}
  {271--280} (\bibinfo {year} {1992})}\BibitemShut {NoStop}%
\bibitem [{\citenamefont {Bigi}\ \emph {et~al.}(1992)\citenamefont {Bigi},
  \citenamefont {Uraltsev},\ and\ \citenamefont {Vainshtein}}]{Bigi:1992su}%
  \BibitemOpen
  \bibfield  {author} {\bibinfo {author} {\bibfnamefont {Ikaros I.~Y.}\
  \bibnamefont {Bigi}}, \bibinfo {author} {\bibfnamefont {N.~G.}\ \bibnamefont
  {Uraltsev}}, \ and\ \bibinfo {author} {\bibfnamefont {A.~I.}\ \bibnamefont
  {Vainshtein}},\ }\bibfield  {title} {\enquote {\bibinfo {title}
  {{Nonperturbative corrections to inclusive beauty and charm decays: QCD
  versus phenomenological models}},}\ }\href {\doibase
  10.1016/0370-2693(92)90908-M} {\bibfield  {journal} {\bibinfo  {journal}
  {Phys. Lett. B}\ }\textbf {\bibinfo {volume} {293}},\ \bibinfo {pages}
  {430--436} (\bibinfo {year} {1992})},\ \bibinfo {note} {[Erratum: Phys.Lett.B
  297, 477--477 (1992)]},\ \Eprint {http://arxiv.org/abs/hep-ph/9207214}
  {arXiv:hep-ph/9207214} \BibitemShut {NoStop}%
\bibitem [{\citenamefont {Bigi}\ \emph {et~al.}(1993)\citenamefont {Bigi},
  \citenamefont {Shifman}, \citenamefont {Uraltsev},\ and\ \citenamefont
  {Vainshtein}}]{Bigi:1993fe}%
  \BibitemOpen
  \bibfield  {author} {\bibinfo {author} {\bibfnamefont {Ikaros I.~Y.}\
  \bibnamefont {Bigi}}, \bibinfo {author} {\bibfnamefont {Mikhail~A.}\
  \bibnamefont {Shifman}}, \bibinfo {author} {\bibfnamefont {N.~G.}\
  \bibnamefont {Uraltsev}}, \ and\ \bibinfo {author} {\bibfnamefont
  {Arkady~I.}\ \bibnamefont {Vainshtein}},\ }\bibfield  {title} {\enquote
  {\bibinfo {title} {{QCD predictions for lepton spectra in inclusive heavy
  flavor decays}},}\ }\href {\doibase 10.1103/PhysRevLett.71.496} {\bibfield
  {journal} {\bibinfo  {journal} {Phys. Rev. Lett.}\ }\textbf {\bibinfo
  {volume} {71}},\ \bibinfo {pages} {496--499} (\bibinfo {year} {1993})},\
  \Eprint {http://arxiv.org/abs/hep-ph/9304225} {arXiv:hep-ph/9304225}
  \BibitemShut {NoStop}%
\bibitem [{\citenamefont {Blok}\ \emph {et~al.}(1994)\citenamefont {Blok},
  \citenamefont {Koyrakh}, \citenamefont {Shifman},\ and\ \citenamefont
  {Vainshtein}}]{Blok:1993va}%
  \BibitemOpen
  \bibfield  {author} {\bibinfo {author} {\bibfnamefont {B.}~\bibnamefont
  {Blok}}, \bibinfo {author} {\bibfnamefont {L.}~\bibnamefont {Koyrakh}},
  \bibinfo {author} {\bibfnamefont {Mikhail~A.}\ \bibnamefont {Shifman}}, \
  and\ \bibinfo {author} {\bibfnamefont {A.~I.}\ \bibnamefont {Vainshtein}},\
  }\bibfield  {title} {\enquote {\bibinfo {title} {{Differential distributions
  in semileptonic decays of the heavy flavors in QCD}},}\ }\href {\doibase
  10.1103/PhysRevD.50.3572} {\bibfield  {journal} {\bibinfo  {journal} {Phys.
  Rev. D}\ }\textbf {\bibinfo {volume} {49}},\ \bibinfo {pages} {3356}
  (\bibinfo {year} {1994})},\ \bibinfo {note} {[Erratum: Phys.Rev.D 50, 3572
  (1994)]},\ \Eprint {http://arxiv.org/abs/hep-ph/9307247}
  {arXiv:hep-ph/9307247} \BibitemShut {NoStop}%
\bibitem [{\citenamefont {Manohar}\ and\ \citenamefont
  {Wise}(1994)}]{Manohar:1993qn}%
  \BibitemOpen
  \bibfield  {author} {\bibinfo {author} {\bibfnamefont {Aneesh~V.}\
  \bibnamefont {Manohar}}\ and\ \bibinfo {author} {\bibfnamefont {Mark~B.}\
  \bibnamefont {Wise}},\ }\bibfield  {title} {\enquote {\bibinfo {title}
  {{Inclusive semileptonic B and polarized Lambda(b) decays from QCD}},}\
  }\href {\doibase 10.1103/PhysRevD.49.1310} {\bibfield  {journal} {\bibinfo
  {journal} {Phys. Rev. D}\ }\textbf {\bibinfo {volume} {49}},\ \bibinfo
  {pages} {1310--1329} (\bibinfo {year} {1994})},\ \Eprint
  {http://arxiv.org/abs/hep-ph/9308246} {arXiv:hep-ph/9308246} \BibitemShut
  {NoStop}%
\bibitem [{\citenamefont {Lenz}(2015)}]{Lenz:2014jha}%
  \BibitemOpen
  \bibfield  {author} {\bibinfo {author} {\bibfnamefont {Alexander}\
  \bibnamefont {Lenz}},\ }\bibfield  {title} {\enquote {\bibinfo {title}
  {{Lifetimes and heavy quark expansion}},}\ }\href {\doibase
  10.1142/S0217751X15430058} {\bibfield  {journal} {\bibinfo  {journal} {Int.
  J. Mod. Phys. A}\ }\textbf {\bibinfo {volume} {30}},\ \bibinfo {pages}
  {1543005} (\bibinfo {year} {2015})},\ \Eprint
  {http://arxiv.org/abs/1405.3601} {arXiv:1405.3601 [hep-ph]} \BibitemShut
  {NoStop}%
\bibitem [{\citenamefont {Beneke}\ \emph {et~al.}(1999)\citenamefont {Beneke},
  \citenamefont {Buchalla}, \citenamefont {Greub}, \citenamefont {Lenz},\ and\
  \citenamefont {Nierste}}]{Beneke:1998sy}%
  \BibitemOpen
  \bibfield  {author} {\bibinfo {author} {\bibfnamefont {M.}~\bibnamefont
  {Beneke}}, \bibinfo {author} {\bibfnamefont {G.}~\bibnamefont {Buchalla}},
  \bibinfo {author} {\bibfnamefont {C.}~\bibnamefont {Greub}}, \bibinfo
  {author} {\bibfnamefont {A.}~\bibnamefont {Lenz}}, \ and\ \bibinfo {author}
  {\bibfnamefont {U.}~\bibnamefont {Nierste}},\ }\bibfield  {title} {\enquote
  {\bibinfo {title} {{Next-to-leading order QCD corrections to the lifetime
  difference of B(s) mesons}},}\ }\href {\doibase
  10.1016/S0370-2693(99)00684-X} {\bibfield  {journal} {\bibinfo  {journal}
  {Phys. Lett. B}\ }\textbf {\bibinfo {volume} {459}},\ \bibinfo {pages}
  {631--640} (\bibinfo {year} {1999})},\ \Eprint
  {http://arxiv.org/abs/hep-ph/9808385} {arXiv:hep-ph/9808385} \BibitemShut
  {NoStop}%
\bibitem [{\citenamefont {Ciuchini}\ \emph {et~al.}(2003)\citenamefont
  {Ciuchini}, \citenamefont {Franco}, \citenamefont {Lubicz}, \citenamefont
  {Mescia},\ and\ \citenamefont {Tarantino}}]{Ciuchini:2003ww}%
  \BibitemOpen
  \bibfield  {author} {\bibinfo {author} {\bibfnamefont {M.}~\bibnamefont
  {Ciuchini}}, \bibinfo {author} {\bibfnamefont {E.}~\bibnamefont {Franco}},
  \bibinfo {author} {\bibfnamefont {V.}~\bibnamefont {Lubicz}}, \bibinfo
  {author} {\bibfnamefont {F.}~\bibnamefont {Mescia}}, \ and\ \bibinfo {author}
  {\bibfnamefont {C.}~\bibnamefont {Tarantino}},\ }\bibfield  {title} {\enquote
  {\bibinfo {title} {{Lifetime differences and CP violation parameters of
  neutral B mesons at the next-to-leading order in QCD}},}\ }\href {\doibase
  10.1088/1126-6708/2003/08/031} {\bibfield  {journal} {\bibinfo  {journal}
  {JHEP}\ }\textbf {\bibinfo {volume} {08}},\ \bibinfo {pages} {031} (\bibinfo
  {year} {2003})},\ \Eprint {http://arxiv.org/abs/hep-ph/0308029}
  {arXiv:hep-ph/0308029} \BibitemShut {NoStop}%
\bibitem [{\citenamefont {Beneke}\ \emph {et~al.}(2003)\citenamefont {Beneke},
  \citenamefont {Buchalla}, \citenamefont {Lenz},\ and\ \citenamefont
  {Nierste}}]{Beneke:2003az}%
  \BibitemOpen
  \bibfield  {author} {\bibinfo {author} {\bibfnamefont {Martin}\ \bibnamefont
  {Beneke}}, \bibinfo {author} {\bibfnamefont {Gerhard}\ \bibnamefont
  {Buchalla}}, \bibinfo {author} {\bibfnamefont {Alexander}\ \bibnamefont
  {Lenz}}, \ and\ \bibinfo {author} {\bibfnamefont {Ulrich}\ \bibnamefont
  {Nierste}},\ }\bibfield  {title} {\enquote {\bibinfo {title} {{CP asymmetry
  in flavor specific B decays beyond leading logarithms}},}\ }\href {\doibase
  10.1016/j.physletb.2003.09.089} {\bibfield  {journal} {\bibinfo  {journal}
  {Phys. Lett. B}\ }\textbf {\bibinfo {volume} {576}},\ \bibinfo {pages}
  {173--183} (\bibinfo {year} {2003})},\ \Eprint
  {http://arxiv.org/abs/hep-ph/0307344} {arXiv:hep-ph/0307344} \BibitemShut
  {NoStop}%
\bibitem [{\citenamefont {Lenz}\ and\ \citenamefont
  {Nierste}(2007)}]{Lenz:2006hd}%
  \BibitemOpen
  \bibfield  {author} {\bibinfo {author} {\bibfnamefont {Alexander}\
  \bibnamefont {Lenz}}\ and\ \bibinfo {author} {\bibfnamefont {Ulrich}\
  \bibnamefont {Nierste}},\ }\bibfield  {title} {\enquote {\bibinfo {title}
  {{Theoretical update of $B_s - \bar{B}_s$ mixing}},}\ }\href {\doibase
  10.1088/1126-6708/2007/06/072} {\bibfield  {journal} {\bibinfo  {journal}
  {JHEP}\ }\textbf {\bibinfo {volume} {06}},\ \bibinfo {pages} {072} (\bibinfo
  {year} {2007})},\ \Eprint {http://arxiv.org/abs/hep-ph/0612167}
  {arXiv:hep-ph/0612167} \BibitemShut {NoStop}%
\bibitem [{\citenamefont {Asatrian}\ \emph {et~al.}(2017)\citenamefont
  {Asatrian}, \citenamefont {Hovhannisyan}, \citenamefont {Nierste},\ and\
  \citenamefont {Yeghiazaryan}}]{Asatrian:2017qaz}%
  \BibitemOpen
  \bibfield  {author} {\bibinfo {author} {\bibfnamefont {H.~M.}\ \bibnamefont
  {Asatrian}}, \bibinfo {author} {\bibfnamefont {Artyom}\ \bibnamefont
  {Hovhannisyan}}, \bibinfo {author} {\bibfnamefont {Ulrich}\ \bibnamefont
  {Nierste}}, \ and\ \bibinfo {author} {\bibfnamefont {Arsen}\ \bibnamefont
  {Yeghiazaryan}},\ }\bibfield  {title} {\enquote {\bibinfo {title} {{Towards
  next-to-next-to-leading-log accuracy for the width difference in the
  $B_s-\bar{B}_s$ system: fermionic contributions to order $(m_c/m_b)^0$ and
  $(m_c/m_b)^1$}},}\ }\href {\doibase 10.1007/JHEP10(2017)191} {\bibfield
  {journal} {\bibinfo  {journal} {JHEP}\ }\textbf {\bibinfo {volume} {10}},\
  \bibinfo {pages} {191} (\bibinfo {year} {2017})},\ \Eprint
  {http://arxiv.org/abs/1709.02160} {arXiv:1709.02160 [hep-ph]} \BibitemShut
  {NoStop}%
\bibitem [{\citenamefont {Asatrian}\ \emph {et~al.}(2020)\citenamefont
  {Asatrian}, \citenamefont {Asatryan}, \citenamefont {Hovhannisyan},
  \citenamefont {Nierste}, \citenamefont {Tumasyan},\ and\ \citenamefont
  {Yeghiazaryan}}]{Asatrian:2020zxa}%
  \BibitemOpen
  \bibfield  {author} {\bibinfo {author} {\bibfnamefont {Hrachia~M.}\
  \bibnamefont {Asatrian}}, \bibinfo {author} {\bibfnamefont {Hrachya~H.}\
  \bibnamefont {Asatryan}}, \bibinfo {author} {\bibfnamefont {Artyom}\
  \bibnamefont {Hovhannisyan}}, \bibinfo {author} {\bibfnamefont {Ulrich}\
  \bibnamefont {Nierste}}, \bibinfo {author} {\bibfnamefont {Sergey}\
  \bibnamefont {Tumasyan}}, \ and\ \bibinfo {author} {\bibfnamefont {Arsen}\
  \bibnamefont {Yeghiazaryan}},\ }\bibfield  {title} {\enquote {\bibinfo
  {title} {{Penguin contribution to the width difference and $CP$ asymmetry in
  $B_q$-$\bar B_q$ mixing at order $\alpha_s^2 N_f$}},}\ }\href {\doibase
  10.1103/PhysRevD.102.033007} {\bibfield  {journal} {\bibinfo  {journal}
  {Phys. Rev. D}\ }\textbf {\bibinfo {volume} {102}},\ \bibinfo {pages}
  {033007} (\bibinfo {year} {2020})},\ \Eprint
  {http://arxiv.org/abs/2006.13227} {arXiv:2006.13227 [hep-ph]} \BibitemShut
  {NoStop}%
\bibitem [{\citenamefont {Hovhannisyan}\ and\ \citenamefont
  {Nierste}(2022)}]{Hovhannisyan:2022miy}%
  \BibitemOpen
  \bibfield  {author} {\bibinfo {author} {\bibfnamefont {Artyom}\ \bibnamefont
  {Hovhannisyan}}\ and\ \bibinfo {author} {\bibfnamefont {Ulrich}\ \bibnamefont
  {Nierste}},\ }\bibfield  {title} {\enquote {\bibinfo {title} {{Addendum to:
  Towards next-to-next-to-leading-log accuracy for the width difference in the
  $\mathbf{B_s-\bar{B}_s}$ system: fermionic contributions to order
  $\mathbf{(m_c/m_b)^0}$ and $\mathbf{(m_c/m_b)^1}$}},}\ }\href@noop {} {\
  (\bibinfo {year} {2022})},\ \Eprint {http://arxiv.org/abs/2204.11907}
  {arXiv:2204.11907 [hep-ph]} \BibitemShut {NoStop}%
\bibitem [{\citenamefont {Gerlach}\ \emph {et~al.}(2021)\citenamefont
  {Gerlach}, \citenamefont {Nierste}, \citenamefont {Shtabovenko},\ and\
  \citenamefont {Steinhauser}}]{Gerlach:2021xtb}%
  \BibitemOpen
  \bibfield  {author} {\bibinfo {author} {\bibfnamefont {Marvin}\ \bibnamefont
  {Gerlach}}, \bibinfo {author} {\bibfnamefont {Ulrich}\ \bibnamefont
  {Nierste}}, \bibinfo {author} {\bibfnamefont {Vladyslav}\ \bibnamefont
  {Shtabovenko}}, \ and\ \bibinfo {author} {\bibfnamefont {Matthias}\
  \bibnamefont {Steinhauser}},\ }\bibfield  {title} {\enquote {\bibinfo {title}
  {{Two-loop QCD penguin contribution to the width difference in
  B$_{s}$\ensuremath{-}$ {\overline{B}}_s $ mixing}},}\ }\href {\doibase
  10.1007/JHEP07(2021)043} {\bibfield  {journal} {\bibinfo  {journal} {JHEP}\
  }\textbf {\bibinfo {volume} {07}},\ \bibinfo {pages} {043} (\bibinfo {year}
  {2021})},\ \Eprint {http://arxiv.org/abs/2106.05979} {arXiv:2106.05979
  [hep-ph]} \BibitemShut {NoStop}%
\bibitem [{\citenamefont {Gerlach}\ \emph
  {et~al.}(2022{\natexlab{a}})\citenamefont {Gerlach}, \citenamefont {Nierste},
  \citenamefont {Shtabovenko},\ and\ \citenamefont
  {Steinhauser}}]{Gerlach:2022wgb}%
  \BibitemOpen
  \bibfield  {author} {\bibinfo {author} {\bibfnamefont {Marvin}\ \bibnamefont
  {Gerlach}}, \bibinfo {author} {\bibfnamefont {Ulrich}\ \bibnamefont
  {Nierste}}, \bibinfo {author} {\bibfnamefont {Vladyslav}\ \bibnamefont
  {Shtabovenko}}, \ and\ \bibinfo {author} {\bibfnamefont {Matthias}\
  \bibnamefont {Steinhauser}},\ }\bibfield  {title} {\enquote {\bibinfo {title}
  {{The width difference in $B - \bar B$ mixing at order $\alpha_s$ and
  beyond}},}\ }\href {\doibase 10.1007/JHEP04(2022)006} {\bibfield  {journal}
  {\bibinfo  {journal} {JHEP}\ }\textbf {\bibinfo {volume} {04}},\ \bibinfo
  {pages} {006} (\bibinfo {year} {2022}{\natexlab{a}})},\ \Eprint
  {http://arxiv.org/abs/2202.12305} {arXiv:2202.12305 [hep-ph]} \BibitemShut
  {NoStop}%
\bibitem [{\citenamefont {Dowdall}\ \emph {et~al.}(2019)\citenamefont
  {Dowdall}, \citenamefont {Davies}, \citenamefont {Horgan}, \citenamefont
  {Lepage}, \citenamefont {Monahan}, \citenamefont {Shigemitsu},\ and\
  \citenamefont {Wingate}}]{Dowdall:2019bea}%
  \BibitemOpen
  \bibfield  {author} {\bibinfo {author} {\bibfnamefont {R.~J.}\ \bibnamefont
  {Dowdall}}, \bibinfo {author} {\bibfnamefont {C.~T.~H.}\ \bibnamefont
  {Davies}}, \bibinfo {author} {\bibfnamefont {R.~R.}\ \bibnamefont {Horgan}},
  \bibinfo {author} {\bibfnamefont {G.~P.}\ \bibnamefont {Lepage}}, \bibinfo
  {author} {\bibfnamefont {C.~J.}\ \bibnamefont {Monahan}}, \bibinfo {author}
  {\bibfnamefont {J.}~\bibnamefont {Shigemitsu}}, \ and\ \bibinfo {author}
  {\bibfnamefont {M.}~\bibnamefont {Wingate}},\ }\bibfield  {title} {\enquote
  {\bibinfo {title} {{Neutral B-meson mixing from full lattice QCD at the
  physical point}},}\ }\href {\doibase 10.1103/PhysRevD.100.094508} {\bibfield
  {journal} {\bibinfo  {journal} {Phys. Rev. D}\ }\textbf {\bibinfo {volume}
  {100}},\ \bibinfo {pages} {094508} (\bibinfo {year} {2019})},\ \Eprint
  {http://arxiv.org/abs/1907.01025} {arXiv:1907.01025 [hep-lat]} \BibitemShut
  {NoStop}%
\bibitem [{\citenamefont {Kirk}\ \emph {et~al.}(2017)\citenamefont {Kirk},
  \citenamefont {Lenz},\ and\ \citenamefont {Rauh}}]{Kirk:2017juj}%
  \BibitemOpen
  \bibfield  {author} {\bibinfo {author} {\bibfnamefont {M.}~\bibnamefont
  {Kirk}}, \bibinfo {author} {\bibfnamefont {A.}~\bibnamefont {Lenz}}, \ and\
  \bibinfo {author} {\bibfnamefont {T.}~\bibnamefont {Rauh}},\ }\bibfield
  {title} {\enquote {\bibinfo {title} {{Dimension-six matrix elements for meson
  mixing and lifetimes from sum rules}},}\ }\href {\doibase
  10.1007/JHEP12(2017)068} {\bibfield  {journal} {\bibinfo  {journal} {JHEP}\
  }\textbf {\bibinfo {volume} {12}},\ \bibinfo {pages} {068} (\bibinfo {year}
  {2017})},\ \bibinfo {note} {[Erratum: JHEP 06, 162 (2020)]},\ \Eprint
  {http://arxiv.org/abs/1711.02100} {arXiv:1711.02100 [hep-ph]} \BibitemShut
  {NoStop}%
\bibitem [{\citenamefont {King}\ \emph {et~al.}(2021)\citenamefont {King},
  \citenamefont {Lenz},\ and\ \citenamefont {Rauh}}]{King:2021jsq}%
  \BibitemOpen
  \bibfield  {author} {\bibinfo {author} {\bibfnamefont {Daniel}\ \bibnamefont
  {King}}, \bibinfo {author} {\bibfnamefont {Alexander}\ \bibnamefont {Lenz}},
  \ and\ \bibinfo {author} {\bibfnamefont {Thomas}\ \bibnamefont {Rauh}},\
  }\bibfield  {title} {\enquote {\bibinfo {title} {{$SU(3)$ breaking effects in
  $B$ and $D$ meson lifetimes}},}\ }\href@noop {} {\  (\bibinfo {year}
  {2021})},\ \Eprint {http://arxiv.org/abs/2112.03691} {arXiv:2112.03691
  [hep-ph]} \BibitemShut {NoStop}%
\bibitem [{\citenamefont {Beneke}\ \emph {et~al.}(1996)\citenamefont {Beneke},
  \citenamefont {Buchalla},\ and\ \citenamefont {Dunietz}}]{Beneke:1996gn}%
  \BibitemOpen
  \bibfield  {author} {\bibinfo {author} {\bibfnamefont {M.}~\bibnamefont
  {Beneke}}, \bibinfo {author} {\bibfnamefont {G.}~\bibnamefont {Buchalla}}, \
  and\ \bibinfo {author} {\bibfnamefont {I.}~\bibnamefont {Dunietz}},\
  }\bibfield  {title} {\enquote {\bibinfo {title} {{Width Difference in the
  $B_s-\bar{B_s}$ System}},}\ }\href {\doibase 10.1103/PhysRevD.54.4419}
  {\bibfield  {journal} {\bibinfo  {journal} {Phys. Rev. D}\ }\textbf {\bibinfo
  {volume} {54}},\ \bibinfo {pages} {4419--4431} (\bibinfo {year} {1996})},\
  \bibinfo {note} {[Erratum: Phys.Rev.D 83, 119902 (2011)]},\ \Eprint
  {http://arxiv.org/abs/hep-ph/9605259} {arXiv:hep-ph/9605259} \BibitemShut
  {NoStop}%
\bibitem [{\citenamefont {Davies}\ \emph {et~al.}(2020)\citenamefont {Davies},
  \citenamefont {Harrison}, \citenamefont {Lepage}, \citenamefont {Monahan},
  \citenamefont {Shigemitsu},\ and\ \citenamefont {Wingate}}]{Davies:2019gnp}%
  \BibitemOpen
  \bibfield  {author} {\bibinfo {author} {\bibfnamefont {Christine T.~H.}\
  \bibnamefont {Davies}}, \bibinfo {author} {\bibfnamefont {Judd}\ \bibnamefont
  {Harrison}}, \bibinfo {author} {\bibfnamefont {G.~Peter}\ \bibnamefont
  {Lepage}}, \bibinfo {author} {\bibfnamefont {Christopher~J.}\ \bibnamefont
  {Monahan}}, \bibinfo {author} {\bibfnamefont {Junko}\ \bibnamefont
  {Shigemitsu}}, \ and\ \bibinfo {author} {\bibfnamefont {Matthew}\
  \bibnamefont {Wingate}} (\bibinfo {collaboration} {HPQCD}),\ }\bibfield
  {title} {\enquote {\bibinfo {title} {{Lattice QCD matrix elements for the
  ${B_s^0-\bar{B}_s^0}$ width difference beyond leading order}},}\ }\href
  {\doibase 10.1103/PhysRevLett.124.082001} {\bibfield  {journal} {\bibinfo
  {journal} {Phys. Rev. Lett.}\ }\textbf {\bibinfo {volume} {124}},\ \bibinfo
  {pages} {082001} (\bibinfo {year} {2020})},\ \Eprint
  {http://arxiv.org/abs/1910.00970} {arXiv:1910.00970 [hep-lat]} \BibitemShut
  {NoStop}%
\bibitem [{\citenamefont {Chetyrkin}\ \emph
  {et~al.}(1998{\natexlab{a}})\citenamefont {Chetyrkin}, \citenamefont
  {Misiak},\ and\ \citenamefont {Munz}}]{Chetyrkin:1997gb}%
  \BibitemOpen
  \bibfield  {author} {\bibinfo {author} {\bibfnamefont {Konstantin~G.}\
  \bibnamefont {Chetyrkin}}, \bibinfo {author} {\bibfnamefont {Mikolaj}\
  \bibnamefont {Misiak}}, \ and\ \bibinfo {author} {\bibfnamefont {Manfred}\
  \bibnamefont {Munz}},\ }\bibfield  {title} {\enquote {\bibinfo {title}
  {{$|\Delta F| = 1$ nonleptonic effective Hamiltonian in a simpler scheme}},}\
  }\href {\doibase 10.1016/S0550-3213(98)00131-X} {\bibfield  {journal}
  {\bibinfo  {journal} {Nucl. Phys. B}\ }\textbf {\bibinfo {volume} {520}},\
  \bibinfo {pages} {279--297} (\bibinfo {year} {1998}{\natexlab{a}})},\ \Eprint
  {http://arxiv.org/abs/hep-ph/9711280} {arXiv:hep-ph/9711280} \BibitemShut
  {NoStop}%
\bibitem [{\citenamefont {Nogueira}(1993)}]{Nogueira:1991ex}%
  \BibitemOpen
  \bibfield  {author} {\bibinfo {author} {\bibfnamefont {Paulo}\ \bibnamefont
  {Nogueira}},\ }\bibfield  {title} {\enquote {\bibinfo {title} {{Automatic
  Feynman graph generation}},}\ }\href {\doibase 10.1006/jcph.1993.1074}
  {\bibfield  {journal} {\bibinfo  {journal} {J. Comput. Phys.}\ }\textbf
  {\bibinfo {volume} {105}},\ \bibinfo {pages} {279--289} (\bibinfo {year}
  {1993})}\BibitemShut {NoStop}%
\bibitem [{\citenamefont {Gerlach}\ \emph
  {et~al.}(2022{\natexlab{b}})\citenamefont {Gerlach}, \citenamefont {Herren},\
  and\ \citenamefont {Lang}}]{Gerlach:2022qnc}%
  \BibitemOpen
  \bibfield  {author} {\bibinfo {author} {\bibfnamefont {Marvin}\ \bibnamefont
  {Gerlach}}, \bibinfo {author} {\bibfnamefont {Florian}\ \bibnamefont
  {Herren}}, \ and\ \bibinfo {author} {\bibfnamefont {Martin}\ \bibnamefont
  {Lang}},\ }\bibfield  {title} {\enquote {\bibinfo {title} {{$\texttt{tapir}$:
  A tool for topologies, amplitudes, partial fraction decomposition and input
  for reductions}},}\ }\href@noop {} {\  (\bibinfo {year}
  {2022}{\natexlab{b}})},\ \Eprint {http://arxiv.org/abs/2201.05618}
  {arXiv:2201.05618 [hep-ph]} \BibitemShut {NoStop}%
\bibitem [{\citenamefont {Harlander}\ \emph {et~al.}(1998)\citenamefont
  {Harlander}, \citenamefont {Seidensticker},\ and\ \citenamefont
  {Steinhauser}}]{Harlander:1998cmq}%
  \BibitemOpen
  \bibfield  {author} {\bibinfo {author} {\bibfnamefont {R.}~\bibnamefont
  {Harlander}}, \bibinfo {author} {\bibfnamefont {T.}~\bibnamefont
  {Seidensticker}}, \ and\ \bibinfo {author} {\bibfnamefont {M.}~\bibnamefont
  {Steinhauser}},\ }\bibfield  {title} {\enquote {\bibinfo {title} {{Complete
  corrections of Order alpha alpha-s to the decay of the Z boson into bottom
  quarks}},}\ }\href {\doibase 10.1016/S0370-2693(98)00220-2} {\bibfield
  {journal} {\bibinfo  {journal} {Phys. Lett. B}\ }\textbf {\bibinfo {volume}
  {426}},\ \bibinfo {pages} {125--132} (\bibinfo {year} {1998})},\ \Eprint
  {http://arxiv.org/abs/hep-ph/9712228} {arXiv:hep-ph/9712228} \BibitemShut
  {NoStop}%
\bibitem [{\citenamefont {Seidensticker}(1999)}]{Seidensticker:1999bb}%
  \BibitemOpen
  \bibfield  {author} {\bibinfo {author} {\bibfnamefont {T.}~\bibnamefont
  {Seidensticker}},\ }\bibfield  {title} {\enquote {\bibinfo {title}
  {{Automatic application of successive asymptotic expansions of Feynman
  diagrams}},}\ }in\ \href@noop {} {\emph {\bibinfo {booktitle} {{6th
  International Workshop on New Computing Techniques in Physics Research:
  Software Engineering, Artificial Intelligence Neural Nets, Genetic
  Algorithms, Symbolic Algebra, Automatic Calculation}}}}\ (\bibinfo {year}
  {1999})\ \Eprint {http://arxiv.org/abs/hep-ph/9905298} {arXiv:hep-ph/9905298}
  \BibitemShut {NoStop}%
\bibitem [{\citenamefont {Smirnov}\ and\ \citenamefont
  {Chuharev}(2020)}]{Smirnov:2019qkx}%
  \BibitemOpen
  \bibfield  {author} {\bibinfo {author} {\bibfnamefont {A.~V.}\ \bibnamefont
  {Smirnov}}\ and\ \bibinfo {author} {\bibfnamefont {F.~S.}\ \bibnamefont
  {Chuharev}},\ }\bibfield  {title} {\enquote {\bibinfo {title} {{FIRE6:
  Feynman Integral REduction with Modular Arithmetic}},}\ }\href {\doibase
  10.1016/j.cpc.2019.106877} {\bibfield  {journal} {\bibinfo  {journal}
  {Comput. Phys. Commun.}\ }\textbf {\bibinfo {volume} {247Â }},\ \bibinfo
  {pages} {106877} (\bibinfo {year} {2020})},\ \Eprint
  {http://arxiv.org/abs/1901.07808} {arXiv:1901.07808 [hep-ph]} \BibitemShut
  {NoStop}%
\bibitem [{\citenamefont {Lee}(2012)}]{Lee:2012cn}%
  \BibitemOpen
  \bibfield  {author} {\bibinfo {author} {\bibfnamefont {R.~N.}\ \bibnamefont
  {Lee}},\ }\bibfield  {title} {\enquote {\bibinfo {title} {{Presenting
  LiteRed: a tool for the Loop InTEgrals REDuction}},}\ }\href@noop {} {\
  (\bibinfo {year} {2012})},\ \Eprint {http://arxiv.org/abs/1212.2685}
  {arXiv:1212.2685 [hep-ph]} \BibitemShut {NoStop}%
\bibitem [{\citenamefont {Lee}(2014)}]{Lee:2013mka}%
  \BibitemOpen
  \bibfield  {author} {\bibinfo {author} {\bibfnamefont {Roman~N.}\
  \bibnamefont {Lee}},\ }\bibfield  {title} {\enquote {\bibinfo {title}
  {{LiteRed 1.4: a powerful tool for reduction of multiloop integrals}},}\
  }\href {\doibase 10.1088/1742-6596/523/1/012059} {\bibfield  {journal}
  {\bibinfo  {journal} {J. Phys. Conf. Ser.}\ }\textbf {\bibinfo {volume}
  {523}},\ \bibinfo {pages} {012059} (\bibinfo {year} {2014})},\ \Eprint
  {http://arxiv.org/abs/1310.1145} {arXiv:1310.1145 [hep-ph]} \BibitemShut
  {NoStop}%
\bibitem [{\citenamefont {Mertig}\ \emph {et~al.}(1991)\citenamefont {Mertig},
  \citenamefont {Bohm},\ and\ \citenamefont {Denner}}]{Mertig:1990an}%
  \BibitemOpen
  \bibfield  {author} {\bibinfo {author} {\bibfnamefont {R.}~\bibnamefont
  {Mertig}}, \bibinfo {author} {\bibfnamefont {M.}~\bibnamefont {Bohm}}, \ and\
  \bibinfo {author} {\bibfnamefont {Ansgar}\ \bibnamefont {Denner}},\
  }\bibfield  {title} {\enquote {\bibinfo {title} {{FEYN CALC: Computer
  algebraic calculation of Feynman amplitudes}},}\ }\href {\doibase
  10.1016/0010-4655(91)90130-D} {\bibfield  {journal} {\bibinfo  {journal}
  {Comput. Phys. Commun.}\ }\textbf {\bibinfo {volume} {64}},\ \bibinfo {pages}
  {345--359} (\bibinfo {year} {1991})}\BibitemShut {NoStop}%
\bibitem [{\citenamefont {Shtabovenko}\ \emph {et~al.}(2016)\citenamefont
  {Shtabovenko}, \citenamefont {Mertig},\ and\ \citenamefont
  {Orellana}}]{Shtabovenko:2016sxi}%
  \BibitemOpen
  \bibfield  {author} {\bibinfo {author} {\bibfnamefont {Vladyslav}\
  \bibnamefont {Shtabovenko}}, \bibinfo {author} {\bibfnamefont {Rolf}\
  \bibnamefont {Mertig}}, \ and\ \bibinfo {author} {\bibfnamefont {Frederik}\
  \bibnamefont {Orellana}},\ }\bibfield  {title} {\enquote {\bibinfo {title}
  {{New Developments in FeynCalc 9.0}},}\ }\href {\doibase
  10.1016/j.cpc.2016.06.008} {\bibfield  {journal} {\bibinfo  {journal}
  {Comput. Phys. Commun.}\ }\textbf {\bibinfo {volume} {207}},\ \bibinfo
  {pages} {432--444} (\bibinfo {year} {2016})},\ \Eprint
  {http://arxiv.org/abs/1601.01167} {arXiv:1601.01167 [hep-ph]} \BibitemShut
  {NoStop}%
\bibitem [{\citenamefont {Shtabovenko}\ \emph {et~al.}(2020)\citenamefont
  {Shtabovenko}, \citenamefont {Mertig},\ and\ \citenamefont
  {Orellana}}]{Shtabovenko:2020gxv}%
  \BibitemOpen
  \bibfield  {author} {\bibinfo {author} {\bibfnamefont {Vladyslav}\
  \bibnamefont {Shtabovenko}}, \bibinfo {author} {\bibfnamefont {Rolf}\
  \bibnamefont {Mertig}}, \ and\ \bibinfo {author} {\bibfnamefont {Frederik}\
  \bibnamefont {Orellana}},\ }\bibfield  {title} {\enquote {\bibinfo {title}
  {{FeynCalc 9.3: New features and improvements}},}\ }\href {\doibase
  10.1016/j.cpc.2020.107478} {\bibfield  {journal} {\bibinfo  {journal}
  {Comput. Phys. Commun.}\ }\textbf {\bibinfo {volume} {256}},\ \bibinfo
  {pages} {107478} (\bibinfo {year} {2020})},\ \Eprint
  {http://arxiv.org/abs/2001.04407} {arXiv:2001.04407 [hep-ph]} \BibitemShut
  {NoStop}%
\bibitem [{\citenamefont {Shtabovenko}(2021)}]{Shtabovenko:2021hjx}%
  \BibitemOpen
  \bibfield  {author} {\bibinfo {author} {\bibfnamefont {Vladyslav}\
  \bibnamefont {Shtabovenko}},\ }\bibfield  {title} {\enquote {\bibinfo {title}
  {{FeynCalc goes multiloop}},}\ }in\ \href@noop {} {\emph {\bibinfo
  {booktitle} {{20th International Workshop on Advanced Computing and Analysis
  Techniques in Physics Research}: {AI Decoded - Towards Sustainable, Diverse,
  Performant and Effective Scientific Computing}}}}\ (\bibinfo {year} {2021})\
  \Eprint {http://arxiv.org/abs/2112.14132} {arXiv:2112.14132 [hep-ph]}
  \BibitemShut {NoStop}%
\bibitem [{\citenamefont {Panzer}(2015)}]{Panzer:2015ida}%
  \BibitemOpen
  \bibfield  {author} {\bibinfo {author} {\bibfnamefont {Erik}\ \bibnamefont
  {Panzer}},\ }\emph {\bibinfo {title} {{Feynman integrals and
  hyperlogarithms}}},\ \href {\doibase 10.18452/17157} {Ph.D. thesis},\
  \bibinfo  {school} {Humboldt U.} (\bibinfo {year} {2015}),\ \Eprint
  {http://arxiv.org/abs/1506.07243} {arXiv:1506.07243 [math-ph]} \BibitemShut
  {NoStop}%
\bibitem [{\citenamefont {Duhr}\ and\ \citenamefont
  {Dulat}(2019)}]{Duhr:2019tlz}%
  \BibitemOpen
  \bibfield  {author} {\bibinfo {author} {\bibfnamefont {Claude}\ \bibnamefont
  {Duhr}}\ and\ \bibinfo {author} {\bibfnamefont {Falko}\ \bibnamefont
  {Dulat}},\ }\bibfield  {title} {\enquote {\bibinfo {title} {{PolyLogTools
  \textemdash{} polylogs for the masses}},}\ }\href {\doibase
  10.1007/JHEP08(2019)135} {\bibfield  {journal} {\bibinfo  {journal} {JHEP}\
  }\textbf {\bibinfo {volume} {08}},\ \bibinfo {pages} {135} (\bibinfo {year}
  {2019})},\ \Eprint {http://arxiv.org/abs/1904.07279} {arXiv:1904.07279
  [hep-th]} \BibitemShut {NoStop}%
\bibitem [{\citenamefont {Schnetz}()}]{schnetz}%
  \BibitemOpen
  \bibfield  {author} {\bibinfo {author} {\bibfnamefont {Oliver}\ \bibnamefont
  {Schnetz}},\ }\bibfield  {title} {\enquote {\bibinfo {title}
  {{HyperLogProcedures}},}\ }\href@noop {} {\bibinfo  {journal}
  {\url{https://www.math.fau.de/person/oliver-schnetz}}\ }\BibitemShut
  {NoStop}%
\bibitem [{\citenamefont {Chetyrkin}\ \emph
  {et~al.}(1998{\natexlab{b}})\citenamefont {Chetyrkin}, \citenamefont
  {Misiak},\ and\ \citenamefont {Munz}}]{Chetyrkin:1997fm}%
  \BibitemOpen
\bibfield  {journal} {  }\bibfield  {author} {\bibinfo {author} {\bibfnamefont
  {Konstantin~G.}\ \bibnamefont {Chetyrkin}}, \bibinfo {author} {\bibfnamefont
  {Mikolaj}\ \bibnamefont {Misiak}}, \ and\ \bibinfo {author} {\bibfnamefont
  {Manfred}\ \bibnamefont {Munz}},\ }\bibfield  {title} {\enquote {\bibinfo
  {title} {{Beta functions and anomalous dimensions up to three loops}},}\
  }\href {\doibase 10.1016/S0550-3213(98)00122-9} {\bibfield  {journal}
  {\bibinfo  {journal} {Nucl. Phys. B}\ }\textbf {\bibinfo {volume} {518}},\
  \bibinfo {pages} {473--494} (\bibinfo {year} {1998}{\natexlab{b}})},\ \Eprint
  {http://arxiv.org/abs/hep-ph/9711266} {arXiv:hep-ph/9711266} \BibitemShut
  {NoStop}%
\bibitem [{\citenamefont {Ciuchini}\ \emph {et~al.}(2002)\citenamefont
  {Ciuchini}, \citenamefont {Franco}, \citenamefont {Lubicz},\ and\
  \citenamefont {Mescia}}]{Ciuchini:2001vx}%
  \BibitemOpen
  \bibfield  {author} {\bibinfo {author} {\bibfnamefont {Marco}\ \bibnamefont
  {Ciuchini}}, \bibinfo {author} {\bibfnamefont {E.}~\bibnamefont {Franco}},
  \bibinfo {author} {\bibfnamefont {V.}~\bibnamefont {Lubicz}}, \ and\ \bibinfo
  {author} {\bibfnamefont {F.}~\bibnamefont {Mescia}},\ }\bibfield  {title}
  {\enquote {\bibinfo {title} {{Next-to-leading order QCD corrections to
  spectator effects in lifetimes of beauty hadrons}},}\ }\href {\doibase
  10.1016/S0550-3213(02)00006-8} {\bibfield  {journal} {\bibinfo  {journal}
  {Nucl. Phys. B}\ }\textbf {\bibinfo {volume} {625}},\ \bibinfo {pages}
  {211--238} (\bibinfo {year} {2002})},\ \Eprint
  {http://arxiv.org/abs/hep-ph/0110375} {arXiv:hep-ph/0110375} \BibitemShut
  {NoStop}%
\bibitem [{\citenamefont {Herren}\ and\ \citenamefont
  {Steinhauser}(2018)}]{Herren:2017osy}%
  \BibitemOpen
  \bibfield  {author} {\bibinfo {author} {\bibfnamefont {Florian}\ \bibnamefont
  {Herren}}\ and\ \bibinfo {author} {\bibfnamefont {Matthias}\ \bibnamefont
  {Steinhauser}},\ }\bibfield  {title} {\enquote {\bibinfo {title} {{Version 3
  of RunDec and CRunDec}},}\ }\href {\doibase 10.1016/j.cpc.2017.11.014}
  {\bibfield  {journal} {\bibinfo  {journal} {Comput. Phys. Commun.}\ }\textbf
  {\bibinfo {volume} {224}},\ \bibinfo {pages} {333--345} (\bibinfo {year}
  {2018})},\ \Eprint {http://arxiv.org/abs/1703.03751} {arXiv:1703.03751
  [hep-ph]} \BibitemShut {NoStop}%
\bibitem [{\citenamefont {Zyla}\ \emph {et~al.}(2020)\citenamefont {Zyla} \emph
  {et~al.}}]{ParticleDataGroup:2020ssz}%
  \BibitemOpen
  \bibfield  {author} {\bibinfo {author} {\bibfnamefont {P.~A.}\ \bibnamefont
  {Zyla}} \emph {et~al.} (\bibinfo {collaboration} {Particle Data Group}),\
  }\bibfield  {title} {\enquote {\bibinfo {title} {{Review of Particle
  Physics}},}\ }\href {\doibase 10.1093/ptep/ptaa104} {\bibfield  {journal}
  {\bibinfo  {journal} {PTEP}\ }\textbf {\bibinfo {volume} {2020}},\ \bibinfo
  {pages} {083C01} (\bibinfo {year} {2020})}\BibitemShut {NoStop}%
\bibitem [{\citenamefont {Chetyrkin}\ \emph {et~al.}(2017)\citenamefont
  {Chetyrkin}, \citenamefont {Kuhn}, \citenamefont {Maier}, \citenamefont
  {Maierhofer}, \citenamefont {Marquard}, \citenamefont {Steinhauser},\ and\
  \citenamefont {Sturm}}]{Chetyrkin:2017lif}%
  \BibitemOpen
  \bibfield  {author} {\bibinfo {author} {\bibfnamefont {Konstantin~G.}\
  \bibnamefont {Chetyrkin}}, \bibinfo {author} {\bibfnamefont {Johann~H.}\
  \bibnamefont {Kuhn}}, \bibinfo {author} {\bibfnamefont {Andreas}\
  \bibnamefont {Maier}}, \bibinfo {author} {\bibfnamefont {Philipp}\
  \bibnamefont {Maierhofer}}, \bibinfo {author} {\bibfnamefont {Peter}\
  \bibnamefont {Marquard}}, \bibinfo {author} {\bibfnamefont {Matthias}\
  \bibnamefont {Steinhauser}}, \ and\ \bibinfo {author} {\bibfnamefont
  {Christian}\ \bibnamefont {Sturm}},\ }\bibfield  {title} {\enquote {\bibinfo
  {title} {{Addendum to \textquotedblleft{}Charm and bottom quark masses: An
  update\textquotedblright{}}},}\ }\href {\doibase 10.1103/PhysRevD.96.116007}
  {\  (\bibinfo {year} {2017}),\ 10.1103/PhysRevD.96.116007},\ \bibinfo {note}
  {[Addendum: Phys.Rev.D 96, 116007 (2017)]},\ \Eprint
  {http://arxiv.org/abs/1710.04249} {arXiv:1710.04249 [hep-ph]} \BibitemShut
  {NoStop}%
\bibitem [{\citenamefont {Bazavov}\ \emph {et~al.}(2018)\citenamefont {Bazavov}
  \emph {et~al.}}]{Bazavov:2017lyh}%
  \BibitemOpen
  \bibfield  {author} {\bibinfo {author} {\bibfnamefont {A.}~\bibnamefont
  {Bazavov}} \emph {et~al.},\ }\bibfield  {title} {\enquote {\bibinfo {title}
  {{$B$- and $D$-meson leptonic decay constants from four-flavor lattice
  QCD}},}\ }\href {\doibase 10.1103/PhysRevD.98.074512} {\bibfield  {journal}
  {\bibinfo  {journal} {Phys. Rev. D}\ }\textbf {\bibinfo {volume} {98}},\
  \bibinfo {pages} {074512} (\bibinfo {year} {2018})},\ \Eprint
  {http://arxiv.org/abs/1712.09262} {arXiv:1712.09262 [hep-lat]} \BibitemShut
  {NoStop}%
\bibitem [{\citenamefont {Beneke}(1998)}]{Beneke:1998rk}%
  \BibitemOpen
  \bibfield  {author} {\bibinfo {author} {\bibfnamefont {M.}~\bibnamefont
  {Beneke}},\ }\bibfield  {title} {\enquote {\bibinfo {title} {{A Quark mass
  definition adequate for threshold problems}},}\ }\href {\doibase
  10.1016/S0370-2693(98)00741-2} {\bibfield  {journal} {\bibinfo  {journal}
  {Phys. Lett. B}\ }\textbf {\bibinfo {volume} {434}},\ \bibinfo {pages}
  {115--125} (\bibinfo {year} {1998})},\ \Eprint
  {http://arxiv.org/abs/hep-ph/9804241} {arXiv:hep-ph/9804241} \BibitemShut
  {NoStop}%
\bibitem [{\citenamefont {Bigi}\ \emph {et~al.}(1994)\citenamefont {Bigi},
  \citenamefont {Shifman}, \citenamefont {Uraltsev},\ and\ \citenamefont
  {Vainshtein}}]{Bigi:1994em}%
  \BibitemOpen
  \bibfield  {author} {\bibinfo {author} {\bibfnamefont {Ikaros I.~Y.}\
  \bibnamefont {Bigi}}, \bibinfo {author} {\bibfnamefont {Mikhail~A.}\
  \bibnamefont {Shifman}}, \bibinfo {author} {\bibfnamefont {N.~G.}\
  \bibnamefont {Uraltsev}}, \ and\ \bibinfo {author} {\bibfnamefont {A.~I.}\
  \bibnamefont {Vainshtein}},\ }\bibfield  {title} {\enquote {\bibinfo {title}
  {{The Pole mass of the heavy quark. Perturbation theory and beyond}},}\
  }\href {\doibase 10.1103/PhysRevD.50.2234} {\bibfield  {journal} {\bibinfo
  {journal} {Phys. Rev. D}\ }\textbf {\bibinfo {volume} {50}},\ \bibinfo
  {pages} {2234--2246} (\bibinfo {year} {1994})},\ \Eprint
  {http://arxiv.org/abs/hep-ph/9402360} {arXiv:hep-ph/9402360} \BibitemShut
  {NoStop}%
\bibitem [{\citenamefont {Beneke}\ and\ \citenamefont
  {Braun}(1994)}]{Beneke:1994sw}%
  \BibitemOpen
  \bibfield  {author} {\bibinfo {author} {\bibfnamefont {M.}~\bibnamefont
  {Beneke}}\ and\ \bibinfo {author} {\bibfnamefont {Vladimir~M.}\ \bibnamefont
  {Braun}},\ }\bibfield  {title} {\enquote {\bibinfo {title} {{Heavy quark
  effective theory beyond perturbation theory: Renormalons, the pole mass and
  the residual mass term}},}\ }\href {\doibase 10.1016/0550-3213(94)90314-X}
  {\bibfield  {journal} {\bibinfo  {journal} {Nucl. Phys. B}\ }\textbf
  {\bibinfo {volume} {426}},\ \bibinfo {pages} {301--343} (\bibinfo {year}
  {1994})},\ \Eprint {http://arxiv.org/abs/hep-ph/9402364}
  {arXiv:hep-ph/9402364} \BibitemShut {NoStop}%
\bibitem [{\citenamefont {Beneke}(2021)}]{Beneke:2021lkq}%
  \BibitemOpen
  \bibfield  {author} {\bibinfo {author} {\bibfnamefont {Martin}\ \bibnamefont
  {Beneke}},\ }\bibfield  {title} {\enquote {\bibinfo {title} {{Pole mass
  renormalon and its ramifications}},}\ }\href {\doibase
  10.1140/epjs/s11734-021-00268-w} {\bibfield  {journal} {\bibinfo  {journal}
  {Eur. Phys. J. ST}\ }\textbf {\bibinfo {volume} {230}},\ \bibinfo {pages}
  {2565--2579} (\bibinfo {year} {2021})},\ \Eprint
  {http://arxiv.org/abs/2108.04861} {arXiv:2108.04861 [hep-ph]} \BibitemShut
  {NoStop}%
\bibitem [{\citenamefont {Aaij}\ \emph {et~al.}(2022)\citenamefont {Aaij} \emph
  {et~al.}}]{LHCb:2021moh}%
  \BibitemOpen
  \bibfield  {author} {\bibinfo {author} {\bibfnamefont {R.}~\bibnamefont
  {Aaij}} \emph {et~al.} (\bibinfo {collaboration} {LHCb}),\ }\bibfield
  {title} {\enquote {\bibinfo {title} {{Precise determination of the
  $B_{\mathrm{s}}^0$\textendash{}$\overline B_{\mathrm{s}}^0$ oscillation
  frequency}},}\ }\href {\doibase 10.1038/s41567-021-01394-x} {\bibfield
  {journal} {\bibinfo  {journal} {Nature Phys.}\ }\textbf {\bibinfo {volume}
  {18}},\ \bibinfo {pages} {1--5} (\bibinfo {year} {2022})},\ \Eprint
  {http://arxiv.org/abs/2104.04421} {arXiv:2104.04421 [hep-ex]} \BibitemShut
  {NoStop}%
\bibitem [{\citenamefont {Gerlach}\ \emph {et~al.}()\citenamefont {Gerlach},
  \citenamefont {Nierste}, \citenamefont {Shtabovenko},\ and\ \citenamefont
  {Steinhauser}}]{Gerlach_in_prep}%
  \BibitemOpen
  \bibfield  {author} {\bibinfo {author} {\bibfnamefont {Marvin}\ \bibnamefont
  {Gerlach}}, \bibinfo {author} {\bibfnamefont {Ulrich}\ \bibnamefont
  {Nierste}}, \bibinfo {author} {\bibfnamefont {Vladyslav}\ \bibnamefont
  {Shtabovenko}}, \ and\ \bibinfo {author} {\bibfnamefont {Matthias}\
  \bibnamefont {Steinhauser}},\ }\bibfield  {title} {\enquote {\bibinfo {title}
  {{Next-to-next-to-leading order QCD corrections to the $B$-meson mixing}},}\
  }\href@noop {} {\bibinfo  {journal} {in preparation}\ }\BibitemShut {NoStop}%
\bibitem [{\citenamefont {Bordone}\ \emph {et~al.}(2021)\citenamefont
  {Bordone}, \citenamefont {Capdevila},\ and\ \citenamefont
  {Gambino}}]{Bordone:2021oof}%
  \BibitemOpen
\bibfield  {journal} {  }\bibfield  {author} {\bibinfo {author} {\bibfnamefont
  {Marzia}\ \bibnamefont {Bordone}}, \bibinfo {author} {\bibfnamefont {Bernat}\
  \bibnamefont {Capdevila}}, \ and\ \bibinfo {author} {\bibfnamefont {Paolo}\
  \bibnamefont {Gambino}},\ }\bibfield  {title} {\enquote {\bibinfo {title}
  {{Three loop calculations and inclusive Vcb}},}\ }\href {\doibase
  10.1016/j.physletb.2021.136679} {\bibfield  {journal} {\bibinfo  {journal}
  {Phys. Lett. B}\ }\textbf {\bibinfo {volume} {822}},\ \bibinfo {pages}
  {136679} (\bibinfo {year} {2021})},\ \Eprint
  {http://arxiv.org/abs/2107.00604} {arXiv:2107.00604 [hep-ph]} \BibitemShut
  {NoStop}%
\bibitem [{\citenamefont {Aoki}\ \emph {et~al.}(2021)\citenamefont {Aoki} \emph
  {et~al.}}]{Aoki:2021kgd}%
  \BibitemOpen
  \bibfield  {author} {\bibinfo {author} {\bibfnamefont {Y.}~\bibnamefont
  {Aoki}} \emph {et~al.},\ }\bibfield  {title} {\enquote {\bibinfo {title}
  {{FLAG Review 2021}},}\ }\href@noop {} {\  (\bibinfo {year} {2021})},\
  \Eprint {http://arxiv.org/abs/2111.09849} {arXiv:2111.09849 [hep-lat]}
  \BibitemShut {NoStop}%
\bibitem [{\citenamefont {Vermaseren}(1994)}]{Vermaseren:1994je}%
  \BibitemOpen
  \bibfield  {author} {\bibinfo {author} {\bibfnamefont {J.~A.~M.}\
  \bibnamefont {Vermaseren}},\ }\bibfield  {title} {\enquote {\bibinfo {title}
  {{Axodraw}},}\ }\href {\doibase 10.1016/0010-4655(94)90034-5} {\bibfield
  {journal} {\bibinfo  {journal} {Comput. Phys. Commun.}\ }\textbf {\bibinfo
  {volume} {83}},\ \bibinfo {pages} {45--58} (\bibinfo {year}
  {1994})}\BibitemShut {NoStop}%
\bibitem [{\citenamefont {Binosi}\ and\ \citenamefont
  {Theussl}(2004)}]{Binosi:2003yf}%
  \BibitemOpen
  \bibfield  {author} {\bibinfo {author} {\bibfnamefont {D.}~\bibnamefont
  {Binosi}}\ and\ \bibinfo {author} {\bibfnamefont {L.}~\bibnamefont
  {Theussl}},\ }\bibfield  {title} {\enquote {\bibinfo {title} {{JaxoDraw: A
  Graphical user interface for drawing Feynman diagrams}},}\ }\href {\doibase
  10.1016/j.cpc.2004.05.001} {\bibfield  {journal} {\bibinfo  {journal}
  {Comput. Phys. Commun.}\ }\textbf {\bibinfo {volume} {161}},\ \bibinfo
  {pages} {76--86} (\bibinfo {year} {2004})},\ \Eprint
  {http://arxiv.org/abs/hep-ph/0309015} {arXiv:hep-ph/0309015} \BibitemShut
  {NoStop}%
\end{thebibliography}%

%
%

\end{document}